\documentclass[prd,twocolumn,showpacs,preprintnumbers,amsmath,amssymb,superscriptaddress,nofootinbib]{revtex4}
\usepackage{amsmath}
\usepackage{amssymb}
\usepackage{amsthm}
\usepackage{overpic,color}
\usepackage{afterpage}
\usepackage{epsfig}
\usepackage{float}
\usepackage{rotating}
\usepackage{lscape}
\usepackage{enumerate}
\usepackage{xspace}
\usepackage{url}
\usepackage{upgreek}
\usepackage{dcolumn}      %align table columns on decimal point
\usepackage{bm}           %bold math
\usepackage{multirow}
\usepackage{bigstrut}
\usepackage{upgreek}
\usepackage[tight]{units} %tight is the default
\newcommand{\numu}{\mbox{$\nu_{\mu}$}}                   % nu_mu
\newcommand{\nue}{\mbox{$\nu_{e}$}}                      % nu_e
\newcommand{\nutau}{\mbox{$\nu_{\tau}$}}                 % nu_tau
\newcommand{\ket}[1]{\left|#1\right\rangle}
\newcommand{\num}{\numu}

\newcommand{\nus}{\nu_s}
\newcommand{\nut}{\nu_{\tau}}

\newcommand{\Uo}[1]{{|U_{#1}|^2}}
\newcommand{\Ut}[2]{{|U_{#1}|^2|U_{#2}|^2}}

\begin{document}
%\pagewiselinenumbers

\preprint{FERMILAB-PUB-09-650-E, hep-ex 1001.0336}

\title{Search for sterile neutrino mixing in the MINOS long-baseline experiment}         % Enter your title between curly braces

\newcommand{\Cambridge}{Cavendish Laboratory, University of Cambridge, Madingley Road, Cambridge CB3 0HE, United Kingdom}
\newcommand{\FNAL}{Fermi National Accelerator Laboratory, Batavia, Illinois 60510, USA}
\newcommand{\RAL}{Rutherford Appleton Laboratory, Science and Technologies Facilities Council, OX11 0QX, United Kingdom}
\newcommand{\UCL}{Department of Physics and Astronomy, University College London, Gower Street, London WC1E 6BT, United Kingdom}
\newcommand{\Caltech}{Lauritsen Laboratory, California Institute of Technology, Pasadena, California 91125, USA}
\newcommand{\ANL}{Argonne National Laboratory, Argonne, Illinois 60439, USA}
\newcommand{\Athens}{Department of Physics, University of Athens, GR-15771 Athens, Greece}
\newcommand{\NTUAthens}{Department of Physics, National Tech. University of Athens, GR-15780 Athens, Greece}
\newcommand{\Benedictine}{Physics Department, Benedictine University, Lisle, Illinois 60532, USA}
\newcommand{\BNL}{Brookhaven National Laboratory, Upton, New York 11973, USA}
\newcommand{\CdF}{APC -- Universit\'{e} Paris 7 Denis Diderot, 10, rue Alice Domon et L\'{e}onie Duquet, F-75205 Paris Cedex 13, France}
\newcommand{\Cleveland}{Cleveland Clinic, Cleveland, Ohio 44195, USA}
\newcommand{\Delhi}{Department of Physics \& Astrophysics, University of Delhi, Delhi 110007, India}
\newcommand{\GEHealth}{GE Healthcare, Florence South Carolina 29501, USA}
\newcommand{\Harvard}{Department of Physics, Harvard University, Cambridge, Massachusetts 02138, USA}
\newcommand{\HolyCross}{Holy Cross College, Notre Dame, Indiana 46556, USA}
\newcommand{\IIT}{Physics Division, Illinois Institute of Technology, Chicago, Illinois 60616, USA}
\newcommand{\Iowa}{Department of Physics and Astronomy, Iowa State University, Ames, Iowa 50011 USA}
\newcommand{\Indiana}{Indiana University, Bloomington, Indiana 47405, USA}
\newcommand{\ITEP}{High Energy Experimental Physics Department, ITEP, B. Cheremushkinskaya, 25, 117218 Moscow, Russia}
\newcommand{\JMU}{Physics Department, James Madison University, Harrisonburg, Virginia 22807, USA}
\newcommand{\LASL}{Nuclear Nonproliferation Division, Threat Reduction Directorate, Los Alamos National Laboratory, Los Alamos, New Mexico 87545, USA}
\newcommand{\LANL}{Los Alamos National Laboratory, Los Alamos, New Mexico 87545, USA}
\newcommand{\Lebedev}{Nuclear Physics Department, Lebedev Physical Institute, Leninsky Prospect 53, 119991 Moscow, Russia}
\newcommand{\LLL}{Lawrence Livermore National Laboratory, Livermore, California 94550, USA}
\newcommand{\LBL}{Lawrence Berkeley National Laboratory, Berkeley, California 94720, USA}
\newcommand{\MIT}{Lincoln Laboratory, Massachusetts Institute of Technology, Lexington, Massachusetts 02420, USA}
\newcommand{\Minnesota}{University of Minnesota, Minneapolis, Minnesota 55455, USA}
\newcommand{\Crookston}{Math, Science and Technology Department, University of Minnesota -- Crookston, Crookston, Minnesota 56716, USA}
\newcommand{\Duluth}{Department of Physics, University of Minnesota -- Duluth, Duluth, Minnesota 55812, USA}
\newcommand{\Ohio}{Center for Cosmology and Astro Particle Physics, Ohio State University, Columbus, Ohio 43210 USA}
\newcommand{\Otterbein}{Otterbein College, Westerville, Ohio 43081, USA}
\newcommand{\Oxford}{Subdepartment of Particle Physics, University of Oxford, Oxford OX1 3RH, United Kingdom}
\newcommand{\PennState}{Department of Physics, Pennsylvania State University, State College, Pennsylvania 16802, USA}
\newcommand{\PennU}{Department of Physics and Astronomy, University of Pennsylvania, Philadelphia, Pennsylvania 19104, USA}
\newcommand{\Pittsburgh}{Department of Physics and Astronomy, University of Pittsburgh, Pittsburgh, Pennsylvania 15260, USA}
\newcommand{\IHEP}{Institute for High Energy Physics, Protvino, Moscow Region RU-140284, Russia}
\newcommand{\RoyalH}{Physics Department, Royal Holloway, University of London, Egham, Surrey, TW20 0EX, United Kingdom}
\newcommand{\Carolina}{Department of Physics and Astronomy, University of South Carolina, Columbia, South Carolina 29208, USA}
\newcommand{\SLAC}{Stanford Linear Accelerator Center, Stanford, California 94309, USA}
\newcommand{\Stanford}{Department of Physics, Stanford University, Stanford, California 94305, USA}
\newcommand{\StJohnFisher}{Physics Department, St. John Fisher College, Rochester, New York 14618 USA}
\newcommand{\Sussex}{Department of Physics and Astronomy, University of Sussex, Falmer, Brighton BN1 9QH, United Kingdom}
\newcommand{\TexasAM}{Physics Department, Texas A\&M University, College Station, Texas 77843, USA}
\newcommand{\Texas}{Department of Physics, University of Texas at Austin, 1 University Station C1600, Austin, Texas 78712, USA}
\newcommand{\TechX}{Tech-X Corporation, Boulder, Colorado 80303, USA}
\newcommand{\Tufts}{Physics Department, Tufts University, Medford, Massachusetts 02155, USA}
\newcommand{\UNICAMP}{Universidade Estadual de Campinas, IFGW-UNICAMP, CP 6165, 13083-970, Campinas, SP, Brazil}
\newcommand{\USP}{Instituto de F\'{i}sica, Universidade de S\~{a}o Paulo,  CP 66318, 05315-970, S\~{a}o Paulo, SP, Brazil}
\newcommand{\Warsaw}{Department of Physics, Warsaw University, Ho\.{z}a 69, PL-00-681 Warsaw, Poland}
\newcommand{\Washington}{Physics Department, Western Washington University, Bellingham, Washington 98225, USA}
\newcommand{\WandM}{Department of Physics, College of William \& Mary, Williamsburg, Virginia 23187, USA}
\newcommand{\Wisconsin}{Physics Department, University of Wisconsin, Madison, Wisconsin 53706, USA}
\newcommand{\deceased}{Deceased.}

\affiliation{\ANL}
\affiliation{\Athens}
\affiliation{\Benedictine}
\affiliation{\BNL}
\affiliation{\Caltech}
\affiliation{\Cambridge}
\affiliation{\UNICAMP}
%\affiliation{\CdF}
\affiliation{\FNAL}
\affiliation{\Harvard}
\affiliation{\HolyCross}
\affiliation{\IIT}
\affiliation{\Indiana}
\affiliation{\Iowa}
%\affiliation{\IHEP}
%\affiliation{\ITEP}
%\affiliation{\JMU}
\affiliation{\Lebedev}
\affiliation{\LLL}
\affiliation{\UCL}
\affiliation{\Minnesota}
\affiliation{\Duluth}
\affiliation{\Otterbein}
\affiliation{\Oxford}
\affiliation{\Pittsburgh}
\affiliation{\RAL}
\affiliation{\USP}
\affiliation{\Carolina}
\affiliation{\Stanford}
\affiliation{\Sussex}
\affiliation{\TexasAM}
\affiliation{\Texas}
\affiliation{\Tufts}
\affiliation{\Warsaw}
\affiliation{\Washington}
\affiliation{\WandM}
\affiliation{\Wisconsin}

\author{P.~Adamson}
\affiliation{\FNAL}
%\affiliation{\UCL}
%\affiliation{\Sussex}

\author{C.~Andreopoulos}
\affiliation{\RAL}
%\affiliation{\Athens}

%\author{K.~E.~Arms}
%\affiliation{\Minnesota}

%\author{R.~Armstrong}
%\affiliation{\Indiana}

\author{D.~J.~Auty}
\affiliation{\Sussex}

%\author{S.~Avvakumov}
%\affiliation{\Stanford}

\author{D.~S.~Ayres}
\affiliation{\ANL}

\author{C.~Backhouse}
\affiliation{\Oxford}

%\author{B.~Baller}
%\affiliation{\FNAL}

%\author{B.~Barish}
%\affiliation{\Caltech}

\author{P.~D.~Barnes~Jr.}
\affiliation{\LLL}

\author{G.~Barr}
\affiliation{\Oxford}

\author{W.~L.~Barrett}
\affiliation{\Washington}

%\author{E.~Beall}
%\altaffiliation[Now at\ ]{\Cleveland .}
%\affiliation{\ANL}
%\affiliation{\Minnesota}

%\author{B.~R.~Becker}
%\affiliation{\Minnesota}

%\author{A.~Belias}
%\affiliation{\RAL}

%\author{R.~H.~Bernstein}
%\affiliation{\FNAL}

%\author{D.~Bhattacharya}
%\affiliation{\Pittsburgh}

\author{M.~Bishai}
\affiliation{\BNL}

\author{A.~Blake}
\affiliation{\Cambridge}

%\author{B.~Bock}
%\affiliation{\Duluth}

% ...

\collaboration{The MINOS Collaboration}
\noaffiliation

\date{\today}

\begin{abstract}
A search for depletion of the combined flux of active neutrino species over a \unit[735]{km} baseline is reported using neutral-current interaction data recorded by the MINOS detectors in the NuMI neutrino beam.  Such a depletion is not expected according to conventional interpretations of neutrino oscillation data involving the three known neutrino flavors.  A depletion would be a signature of oscillations or decay to postulated noninteracting sterile neutrinos, scenarios not ruled out by existing data.  From an exposure of \unit[$3.18 \times 10^{20}$]{protons on target} in which  neutrinos of energies between $\sim$\unit[500]{MeV} and \unit[120]{GeV} are produced predominantly as $\nu_\mu$, the visible energy spectrum of candidate neutral-current reactions in the MINOS far-detector is reconstructed.  Comparison of this spectrum to that inferred from a similarly selected near-detector sample shows that of the portion of the $\nu_\mu$ flux  observed to disappear in charged-current interaction data, the fraction that could be converting to a sterile state is less than 52\% at 90\% confidence level (C.L.). The hypothesis that active neutrinos mix with a single sterile neutrino via oscillations is tested by fitting the data to various models. In the particular four-neutrino models considered, the mixing angles $\theta_{24}$ and $\theta_{34}$ are constrained to be less than $11^\circ$ and $56^\circ$ at 90\% C.L., respectively. The possibility that active neutrinos may decay to sterile neutrinos is also investigated. Pure neutrino decay without oscillations is ruled out at 5.4 standard deviations. For the scenario in which active neutrinos decay into sterile states concurrently with neutrino oscillations, a lower limit is established for the neutrino decay lifetime $\tau_3/m_3 > \unit[2.1 \times 10^{-12}]{\text{s/eV}}$ at 90\% C.L.
\end{abstract}


\maketitle

\section{Introduction}
Compelling evidence has been presented demonstrating the disappearance of muon and electron neutrinos as they propagate from their production point.  Disappearance of muon neutrinos has been observed from neutrino fluxes originating in the atmosphere~\cite{previous,superksays} and accelerator beams~\cite{Ahn:2006zza,ref:CCPRD}; disappearance of electron neutrinos has been observed with neutrino fluxes from the Sun~\cite{solarprev,Ahmad:2002jz} and terrestrial reactors~\cite{kamland}.  Super-Kamiokande and other atmospheric neutrino experiments were the first to report significant deficits of $\num$ charged-current interactions from neutrinos propagating over baselines larger than several hundred kilometers. The K2K and MINOS experiments have observed the disappearance using accelerator-beam neutrinos propagating over fixed baselines of \unit[250]{km} and \unit[735]{km}, respectively.  All experiments reporting muon-neutrino disappearance favor pure $\num\rightarrow\nut$ oscillations as the explanation for the observed disappearance of $\num$~\cite{superktau1, superktau2, ref:CCPRL, Ahn:2006zza}. However, more exotic scenarios in which one or more sterile neutrinos, $\nus$, mix with the three active neutrino species remain as viable alternatives.

Long-baseline experiments such as MINOS provide an opportunity to test alternative scenarios by comparing the observed neutral-current interaction rates in near and far-detectors.  Since all active neutrinos, $\nue,~\num$,~and~$\nut$,~participate in the neutral-current interaction, this comparison provides a sensitive probe to the existence of processes that deplete the flux of active neutrinos between the two detectors.  If neutrinos only oscillate among the active flavors, the rate of neutral-current interactions at the far site of a long-baseline experiment remains unchanged from the null-oscillation prediction.  However, if another process occurs concurrently with active neutrino oscillations, the rate of neutral-current interactions at the far site may be different.  Two such possibilities have attracted considerable attention and are the focus of the analysis reported here, (i) active neutrinos oscillating to $\nus$, and (ii) neutrino decay in conjunction with oscillations.

The possible existence of one or more sterile neutrinos that do not couple to the electroweak current but do mix with the active flavors has been discussed extensively in the literature~\cite{steriletheory1,Donini:2007yf,steriletheory2}.  The existence of sterile neutrinos would provide new degrees of freedom that could help clarify certain outstanding problems with the neutrino-mass spectrum~\cite{deGouvea:2006gz} and with heavy element nucleosynthesis in supernovae~\cite{supernovae}.  A recent search for neutrino oscillations in a short baseline experiment provides no evidence for transitions that would imply existence of sterile neutrinos~\cite{AguilarArevalo:2007it}.  

The coupling between sterile neutrinos and the active neutrinos would likely involve the third mass eigenstate. Observations by the SNO experiment show the total flux of active neutrinos from the Sun to agree with expectations from solar models~\cite{Ahmad:2002jz}, thereby limiting the potential coupling of the first or second neutrino-mass eigenstates to a sterile neutrino.  Additionally, the Super-Kamiokande experiment strongly disfavors pure $\num\rightarrow\nus$ mixing~\cite{superktau1,superktau2}, but does not exclude an admixture of subdominant $\num\rightarrow\nus$ mixing with the dominant $\num\rightarrow\nut$ mixing. % MINOS performed the first dedicated search at fixed long baseline for $\num$ oscillating to both $\nut$ and $\nus$ and reported that less than 68\% of disappearing muon neutrinos could have oscillated to sterile neutrinos at the 90\% confidence level~\cite{ref:NCPRL}.  The analysis presented here includes a larger exposure and tests specific models incorporating one sterile state in the neutrino mixing matrix.
MINOS has recently carried out the first dedicated search at fixed long-baseline for $\num$ oscillating to both $\nut$ and $\nus$~\cite{ref:NCPRL}.    The analysis presented here uses a larger exposure and extends the earlier analysis by considering specific models in which a sterile neutrino state is incorporated into the neutrino mixing matrix.

%Also included in this paper is a search for decay of neutrinos into a sterile state~\cite{ref:Decay}.  In this scenario active neutrinos decay due to an instability of the third mass eigenstate, $m_{3}$.  That eigenstate decays into undetected final-state particles and causes a depletion of the neutral-current event rate at the far-detector compared to expectations.    MINOS has previously used the comparison of its near-detector and far-detector charged-current energy spectra to disfavor pure decay at a level of 3.7 standard deviations~\cite{ref:CCPRL}.  The analysis reported here is the first to directly test the combination of neutrino decay and oscillations in a long-baseline experiment.

The possibility that a neutrino may decay into a sterile state~\cite{ref:Decay} is also explored in this work.   In this scenario,  the mass eigenstate $\nu_{3}$ is unstable and allows active neutrinos to decay into undetectable final states.   The decays would give rise to an anomalous depletion of the neutral-current event rate observed at the far-detector.  The occurrence of pure neutrino decay, without oscillation, has already been shown by MINOS and Super-Kamiokande to be highly disfavored~\cite{ref:CCPRL,ref:superKdecay}.    The analysis reported here represents the first direct test of the neutrino-decay-with-oscillations scenario in a long-baseline experiment.

\section{NuMI Beam and MINOS Detectors}\label{sec:Detectors}
Neutrinos from the NuMI (Neutrinos from the Main Injector) beam~\cite{ref:nim} originate from decays of pions and kaons produced in the beamline target; a significantly smaller contribution arises from subsequent muon decays. The secondary mesons are created using \unit[120]{GeV} protons extracted from the Fermilab Main Injector  incident on a graphite target. The proton extraction occurs in \unit[10]{$\mu$s} spills with a \unit[2.2]{\,s} cycle. Positioned downstream of the target are two parabolic magnetic horns which focus $\pi^{+}$ and $K^{+}$ secondary particles. The focused mesons proceed into a \unit[675]{m} long evacuated decay pipe, where they may decay into muons and neutrinos. The remnant hadrons are stopped by a beam absorber placed at the end of the decay pipe. The tertiary muons are stopped by \unit[240]{m} of rock between the end of the decay volume and the near-detector cavern so that only neutrinos reach the near-detector.  The neutrino energy spectrum can be changed by adjusting either the horn current or the position of the target relative to the horns.  The data employed in this analysis were obtained using the low-energy beam configuration, in which the peak neutrino energy is \unit[3.3]{GeV}~\cite{ref:CCPRD}, and correspond to a far-detector exposure of \unit[$3.18 \times 10^{20}$]{protons on target}, collected during the period of May~2005 to July~2007. In this configuration, according to Monte Carlo simulations, the neutrino flavor composition of the beam is 91.8\% $\num$, 6.9\% $\overline{\nu}_{\mu}$, and 1.3\% $\nue+\overline{\nu}_{e}$. For this analysis the neutrinos and antineutrinos are assumed to oscillate with the same parameters.

The MINOS detectors are planar steel/scintillator tracking calorimeters~\cite{ref:nim}.  The vertically oriented detector planes are composed of \unit[2.54]{cm} thick steel and \unit[1]{cm} thick plastic scintillator. A scintillator layer is composed of \unit[4.1]{cm} wide strips.  Each strip is coupled via a wavelength-shifting fiber to one pixel of a multianode photomultiplier tube (PMT)~\cite{Tagg:2004bu,Lang:2005xu}.

The MINOS near-detector is located \unit[1.04]{km} downstream of the target, has a total mass of \unit[980]{\text{metric tonnes}}, and lies \unit[103]{m} underground at Fermilab. The detector consists of two sections, a calorimeter encompassing the upstream 121 planes and a spectrometer containing the downstream 161 planes. In both sections, one out of every five planes is fully covered with 96 scintillator strips attached to the steel plates. In the calorimeter section, the other four out of five planes are partially covered with 64 scintillator strips, whereas in the spectrometer section no scintillator is attached to the steel. The far-detector is \unit[734]{km} downstream of the near-detector, has a total mass of \unit[5400]{\text{metric tonnes}}, and is located in the Soudan Underground Laboratory, \unit[705]{m} below the surface. It is composed of 484 fully instrumented planes organized in two supermodules~\cite{ref:CCPRD}.  The fiducial masses used for the near and far-detectors are \unit[27]{\text{metric tonnes}} and \unit[3800]{\text{metric tonnes}} respectively. The near-detector steel is  magnetized with an average field intensity of \unit[1.3]{T} whereas the far-detector has an average field of \unit[1.4]{T} in the steel.

\section{Data Selection}\label{sec:Selection}
A neutrino interacting in one of the MINOS detectors produces either a charged-current event with a charged lepton plus hadrons emerging from the event vertex or a neutral-current event with hadrons but no charged lepton in the final state. In either case, the particles in the final state deposit energy in the scintillator strips, which is converted to light and collected by optical fibers and converted to electronic signals by PMTs. The MINOS reconstruction algorithms use event topology and the recorded time stamps of the strips where energy was deposited to identify neutrino events inside the detector. Events must have at least four strips with signal to be considered in the analysis.  Individual scintillator strips are grouped into either reconstructed tracks or showers, and the tracks and showers are combined into events~\cite{ref:CCPRD}. The vertex of each event is required to be sufficiently far from any edge of the detector to ensure that the final-state hadronic showers are contained within the instrumented regions of the detectors. On average, each GeV of energy deposition in a neutral-current event induces activity in 12 scintillator strips.

The Monte Carlo simulation of the neutrino beam utilizes FLUKA05~\cite{ref:fluka} to model hadroproduction in the NuMI target and a GEANT3~\cite{ref:geant} simulation of the NuMI beam line to propagate the particles exiting the target. The neutrino interactions in the MINOS detectors are modeled by the NEUGEN-v3~\cite{ref:neugen} program. The simulated neutrino flux  is constrained to agree with the neutrino energy spectra measured in the near-detector for nine different beam configurations~\cite{ref:CCPRD}. This procedure reduces the uncertainties due to the neutrino flux in the far-detector prediction.

\subsection{Data integrity}

All of the data accepted for this analysis must pass a series of requirements on the beam and detector performance. The beam is required to strike within \unit[2]{mm} of the center of the upstream face of the NuMI target, a segmented rectangular graphite rod measuring \unit[6.4]{mm} in width, \unit[15]{mm} in height and \unit[940]{mm} in length~\cite{ref:CCPRD}. The full width at half-maximum of the beam at the target is required to be between \unit[0.1]{mm} and \unit[2.0]{mm} in the transverse horizontal direction and between \unit[0.1]{mm} and \unit[1.5]{mm} in the transverse vertical direction. The minimum allowed beam intensity during a beam spill is \unit[$0.5\times 10^{12}$]{protons on target}.

For all data used in this analysis, all detector subsystems that affect data quality are required to be in normal, stable modes of operation. Checks are made to ensure the coil currents that energize the magnetic fields are at their nominal values in both detectors. The timing between the detectors is synchronized using the Global Positioning System to within  \unit[1]{${\mu}$s} to ensure that correct beam extraction timing is provided to the far-detector electronics. The timing synchronization is not affected by the timing resolution of the detectors, which is \unit[18.8]{ns} and \unit[1.9]{ns} for the near and far-detectors, respectively. The high voltage supplied to each PMT is required  to be at its nominal value.

\subsection{Fiducial requirements}
Only those events reconstructed in the fiducial volume are included in the analysis. In both detectors the reconstructed event vertex is required to be more than \unit[50]{cm} from the edges of the instrumented regions and, in the case of the far-detector, more than \unit[45]{cm} from the center of the magnetic coil that runs through the middle of each detector plane. In addition, a longitudinal veto region comprising either 30 planes at the front of the near-detector or four planes at the front of each of the far-detector supermodules eliminates events that enter through the first plane of a detector but may have originated outside the detector volume. To ensure good shower containment, the event vertex is required to be reconstructed more than \unit[1]{m} from the last plane of each far-detector supermodule and more than \unit[1]{m} from the last plane in the calorimeter region of the near-detector. The sparsely instrumented spectrometer region of the near-detector is not included in the fiducial volume.  The primary vertex of an event is assigned according to the event's reconstructed track vertex. However, for events without a reconstructed track, the vertex of the hadronic shower is used as the event vertex.  

\subsection{Near-detector event selection}

The reconstruction algorithms are designed to handle the high rate of interactions occurring in the near-detector during running with typical intensities of \unit[$2.2\times10^{13}$]{protons on target} per beam spill. At this intensity, an average of 16 neutrino interactions occur in the near-detector for each spill.  For the majority of events the algorithms perform very well. However, for certain event subclassess, shortfalls have been identified and quantified using special studies including low intensity beam data and visual scanning. Monte Carlo studies show that 8\% of all reconstructed events classified as neutral-current interactions are assigned a value of recontructed energy that is less than half of the actual deposited energy in the detector for the simulated interaction; the remaining energy has been reconstructed as a separate event, resulting in an overestimate in the number of reconstructed events. In particular, the number of events with visible energy below \unit[1]{GeV} that are classified as neutral-current candidates is overestimated by 34\%. Therefore, these poorly reconstructed events affect the neutral-current energy spectrum and care has been taken in identifying and removing them from the analysis. 

Reconstruction failures are classified into three main categories: (i) {\it split events}, (ii) {\it leakage events}, and (iii) {\it incomplete events}. Split events occur when a single neutrino interaction results in two or more reconstructed events. Leakage events are due to incorrectly assigned event vertices causing neutrino interactions outside the fiducial volume to be reconstructed within it. The incomplete event category is a looser classification that refers to further types of failures in shower reconstruction to be described below. In all three categories, the visible energy of a neutrino candidate may be underestimated, resulting in a background to neutral-current events at low energies. As near-detector data are used to predict the expected spectra at the far-detector, reconstruction failures specific to the near-detector must be minimized. A set of selection requirements was developed~\cite{ref:rauferThesis} to reduce the occurrence of these failure modes.  After applying the requirements, detailed below, simulations show that the background of poorly reconstructed events having visible energy below \unit[1]{GeV} has been reduced to 8\%.

\begin{figure}
\begin{center}
  \includegraphics[width=1.05\linewidth]{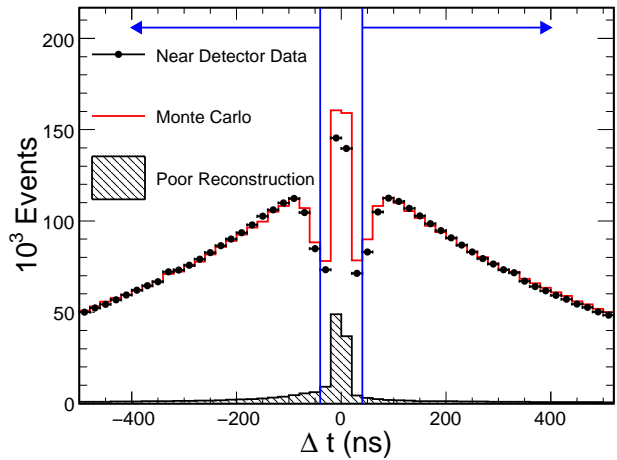}
  \caption{The distribution of time separation between events $\Delta t$ for data (solid points) and Monte Carlo simulated data (solid histogram).
  A background component arising from poorly reconstructed events (hatched histogram) is confined to a narrow region of low $\Delta t$ values. Events accepted for further analysis are indicated by the arrows.}
  \label{fig:splitvariables}
\end{center}
\end{figure}

Split events lead to double-counting of neutrino interactions and incorrect energy reconstruction. If two reconstructed events are caused by the same neutrino interaction, they are expected to appear close in time and in space. The time for each event has been taken to be the median of the recorded signal arrival times for event strips contained within five planes of the event vertex. To minimize the occurrence of split events, the time separation between an event and the closest other event in time is required to be  $|\Delta t| > \unit[40]{ns}$, as shown in Fig.~\ref{fig:splitvariables}. A requirement that the spatial separation between events along the beam direction $|\Delta z|>\unit[1]{m}$ if $\unit[40]{ns} < |\Delta t| < \unit[120]{ns}$ is also employed to further eliminate split events.

%The preselection cuts
%have studied in order to eliminate leaking events that systematically
%have a reconstructed energy lower than the visible energy deposited in
%the detector.  
Leakage events are typically cosmic-ray muons causing steep showers with a high concentration of hit strips in a small number of detector planes. These events can be removed by selecting on this topological characteristic. Thus, a requirement is placed on the ratio of the average number of active strips per plane to the total number of planes with active strips in the event, represented more concisely as (active event strips)/(active event planes)$^2$. Only events for which the ratio is less than 1.0 are accepted by the analysis.

Another type of leakage event is due to secondary particles from interactions occurring outside of the fiducial volume. In the partially covered planes of the near-detector, the steel is instrumented with scintillator to within \unit[16]{cm} from the left-hand-side edges of the steel plate and \unit[1.4]{m} from the right-hand-side edges as viewed from along the beam direction. Consequently, secondary particles may enter the detector laterally due to the sparse instrumentation on the sides. For such cases the reconstruction algorithm is likely to fail in associating hits to events. Nevertheless, the extra activity at the edges of the fully covered planes is recorded and can be used to veto events within a time window. The veto criterion uses the number of active strips and the pulse-height in the edge regions of the detector. An event is accepted if the number of active strips in the veto region recorded within a $\pm$\unit[40]{ns} window around the event vertex time is less than four and the pulse-height deposited in the veto region is less than \unit[2]{MIP}\footnote{Minimum ionizing particle, equivalent to the response produced by a \unit[1]{GeV} muon traversing a detector plane at normal incidence.}. These veto criteria are applied to events with visible energy less than \unit[5]{GeV} in which the number of planes assigned to the reconstructed shower is greater than the number of planes assigned to the reconstructed track, as leakage events are reconstructed as low-energy showers without a clearly defined track.

\begin{figure}
\begin{center}
  \includegraphics[width=1.05\linewidth]{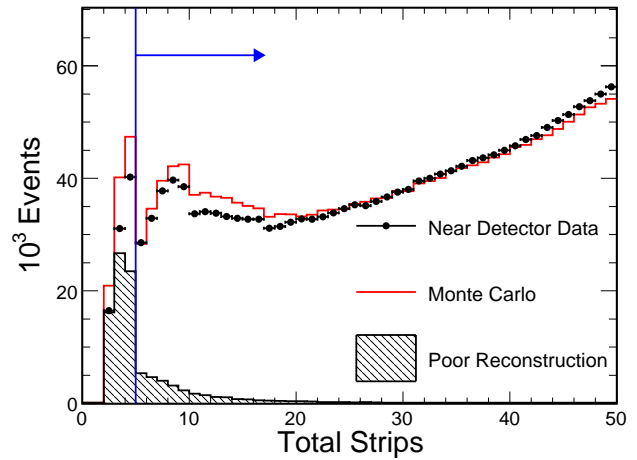}
  \caption{Section of the distribution of the number of strips with nonzero pulse-height, per event, after all other selections are applied. The region displayed, in which the contribution from poorly reconstructed events (hatched histogram) is significant, corresponds to low strip counts and represents a small fraction of the total number of events. The event range accepted for the analysis is identified by the arrow.}
  \label{strips_cleaning}
\end{center}
\end{figure}

Incomplete events arise when the shower reconstruction fails to assign all event-related strips to the shower. This type of reconstruction failure occurs if there are large gaps in a shower or if the shower is generally sparse. In a majority of these cases, events have a very low number of reconstructed strips. Figure~\ref{strips_cleaning} shows the distribution of the number of strips for the events, after applying the selection requirements. To minimize the number of incomplete events in the near-detector data sample, an event is required to have total number of strips greater than four.

In summary, the selection criteria applied to the near-detector data are as follows: (i) the modulus of the time separation between events, $|\Delta t|$, must exceed \unit[40]{ns}; (ii) if \unit[40]{ns}~$<\,|\Delta\,t|\,<$~\unit[120]{ns}, the modulus of the spatial separation between events, $|\Delta z|$, must exceed \unit[1]{m}; (iii) the ratio (active  event strips)/(active event planes)$^2$  must be less than unity; (iv) for events with less than \unit[5]{GeV} of reconstructed energy in which the number of planes is larger in the reconstructed shower than in the reconstructed track, the number of event strips reconstructed  in the detector's veto regions should be less than four and the total pulse-height in those regions must be less than \unit[2]{MIP}; and (v) the total number of strips reconstructed in the event must be more than four. Only events that satisfy these criteria are used for further analysis.

\subsection{Far-detector event selection}

In contrast to the multiple events per beam spill observed in the near-detector, the rate measured in the far-detector is approximately two~events per day within the beam spill times, so the appropriate requirements for event selection are necessarily different. Specifically, the probability that two or more neutrinos produced in the same \unit[10]{$\mu$s} beam spill will interact in the far-detector is negligible. Therefore, if two events are reconstructed in the same spill, the coincidence is due either to a reconstruction failure or else one of the events has a nonbeam origin. Effects of multiple event reconstruction are mitigated by requiring an event to be used in the analysis to contain at least 75\% of the total deposited energy during the beam spill.

The main background in the far-detector results from detector noise arising from the electronics and PMTs or from spontaneous light emission from the scintillator and wavelength-shifting fibers~\cite{ref:sergei}. The noise from the electronics and PMTs is removed by setting an energy threshold in the PMTs. The spontaneous light emission is removed by requiring accepted events to include at least nine strips or at least \unit[10]{MIP} deposited in the detector. Alternatively, events are also accepted if they include more than five strips and deposit more than \unit[5]{MIP} in the detector.

Muons from cosmic rays are a potential source of background events. Given the \unit[0.2]{Hz} cosmic-ray muon rate at the far-detector, the number of cosmic-ray-induced-muons that may potentially coincide with beam spills is comparable to the number of beam-induced neutrino interactions observed. Most cosmic-ray-induced muons are well reconstructed and are efficiently removed by the fiducial requirement. However, the reconstruction algorithms are optimized to handle recorded energy flow in the general direction of the beam. Problem cases can thus arise with very steep cosmic muons, which are removed by requiring the absolute value of the muon direction cosine in the longitudinal direction, $|p_z|/E$, to be higher than 0.4. In some cases the events are so steep that they are reconstructed only as showers and may be removed by using selection variables that describe the transverse and longitudinal shower profiles. The transverse variable is defined by calculating the root-mean-square (rms) value of the shower strip positions, whereas the longitudinal variable is defined as the ratio of active strips per plane to the total number of active planes in the event. Only those showers with a transverse rms value lower than 0.5 and $\rm{(strips/plane)/(event\;planes)}<1$ are accepted for further analysis. Cosmic-ray muons that stop in the detector can mimic beam events if the end of the stopping muon track is interpreted as the vertex and the track is then reconstructed backwards. These events can be identified by performing a linear fit to the timing distribution for strips on a track as a function of the strip longitudinal position. A fit resulting in a negative slope indicates that the event is a downward-going cosmic-ray muon and not a beam neutrino. The sample contamination from cosmic-ray induced muons after these criteria are applied is estimated to be less than 0.1\%~\cite{thesis:Litchfield}.

Another potential background arises from data recorded while the Light Injection calibration system (LI) is flashing during normal data taking. The light injection events are removed with 99.99\% efficiency by using information from a PMT directly connected to the light injection system. By applying additional requirements based on concentrated detector activity, it is estimated that much less than one LI event is accepted in the entire data sample~\cite{thesis:Litchfield}. Furthermore, application of the LI rejection criteria results in no measurable loss of efficiency for beam-neutrino interactions.

%Checks have been performed to garantee that the neutrino events selected with 
%this method represent an unbiased sample all the neutrino interactions in the
%MINOS near-detector.

\section{Event Classification}\label{sec:NCSelection}
After the selection criteria described in the previous section are applied, the analysis proceeds by distinguishing charged-current events from neutral-current events. Distinct event classification procedures are employed for each sample, as described below. The reconstructed neutrino energy spectra for both event classes are used in the fits described in Sec.~\ref{sec:4Flavor} and Sec.~\ref{sec:Decay}. 

The goal of the event classification is to maximize the efficiency and purity of selected samples of neutral-current and charged-current events. Using Monte Carlo event samples, efficiency is defined as the number of true events of one type which are classified as that type,  divided by the total number of true events of that type that pass the criteria described in Sec.~\ref{sec:Selection}. Purity is defined as the ratio of the number of true events of one type selected to the total number of events selected as that type.

To avoid biases, the methods for identifying neutral-current candidate events and procedures employed in predicting the far-detector spectrum, described in Sec.~\ref{sec:Prediction}, were developed and tested using only the near-detector data and Monte Carlo simulation. All analysis procedures were finalized prior to examining data in the far-detector.

 The neutral-current event classification employs several criteria based on reconstruction variables displaying large differences between neutral-current and charged-current events \cite{thesis:Osiecki}. Charged-current events with short or no apparent tracks and poorly reconstructed events are the two main sources of background. The latter is mitigated by employing the various selections described in Sec.~\ref{sec:Selection}.  The classification variables considered are: {\it event length}, expressed as the difference between the first and last active plane in the event; {\it number of tracks} reconstructed in the event; and {\it track extension}, defined as the difference between track length and shower length.

A sample of candidate neutral-current events is obtained by applying specific requirements on the classification variables. Since neutral-current events are typically shorter than charged-current events, events crossing fewer than 60 planes and for which no track is reconstructed are classified as neutral-current. Because neutral-current events are expected to have short or no reconstructed tracks, events crossing fewer than 60 planes that contain a track are classified as neutral-current if the track extends fewer than 5 planes beyond the shower.  The values chosen maximize sensitivity for detection of sterile-neutrino admixture.  Finally, events that are not classified as neutral-current-like are labeled as charged-current-like if they pass the classification procedures described in a previous MINOS publication\footnote{Candidate charged-current events are selected using a likelihood-based particle identification parameter constructed from three probability density functions. The functions  represent distributions for the variables (i) event length, (ii) fraction of the total event signal in the reconstructed track, and (iii) average signal per plane induced by the track. Further details are presented in Ref.~\cite{ref:CCPRD}.}. These requirements are applied to both near and far-detectors to obtain neutral-current and charged-current event samples. 

The 6\% of near-detector events and 3.5\% of far-detector events that are not classified as either neutral-current or charged-current are not used in further analysis. Evaluation of these removed samples using simulated data shows that approximately 75\% of the events in each detector are charged-current interactions, and the reconstructed energies of those neutrinos distribute in accord with the expectation based on the simulation. The remaining events are neutral-current interactions whose distributions in reconstructed energy are very similar in the two detectors; the latter distributions span the full visible energy spectrum, but with modest accentuation of the lower energy region $\unit[0]{GeV} < E_{\rm reco} < \unit[4]{GeV}$. Therefore, the removal of these events does not introduce analysis biases.

\begin{figure}
\begin{center}
\begin{overpic}[width=1.05\linewidth]{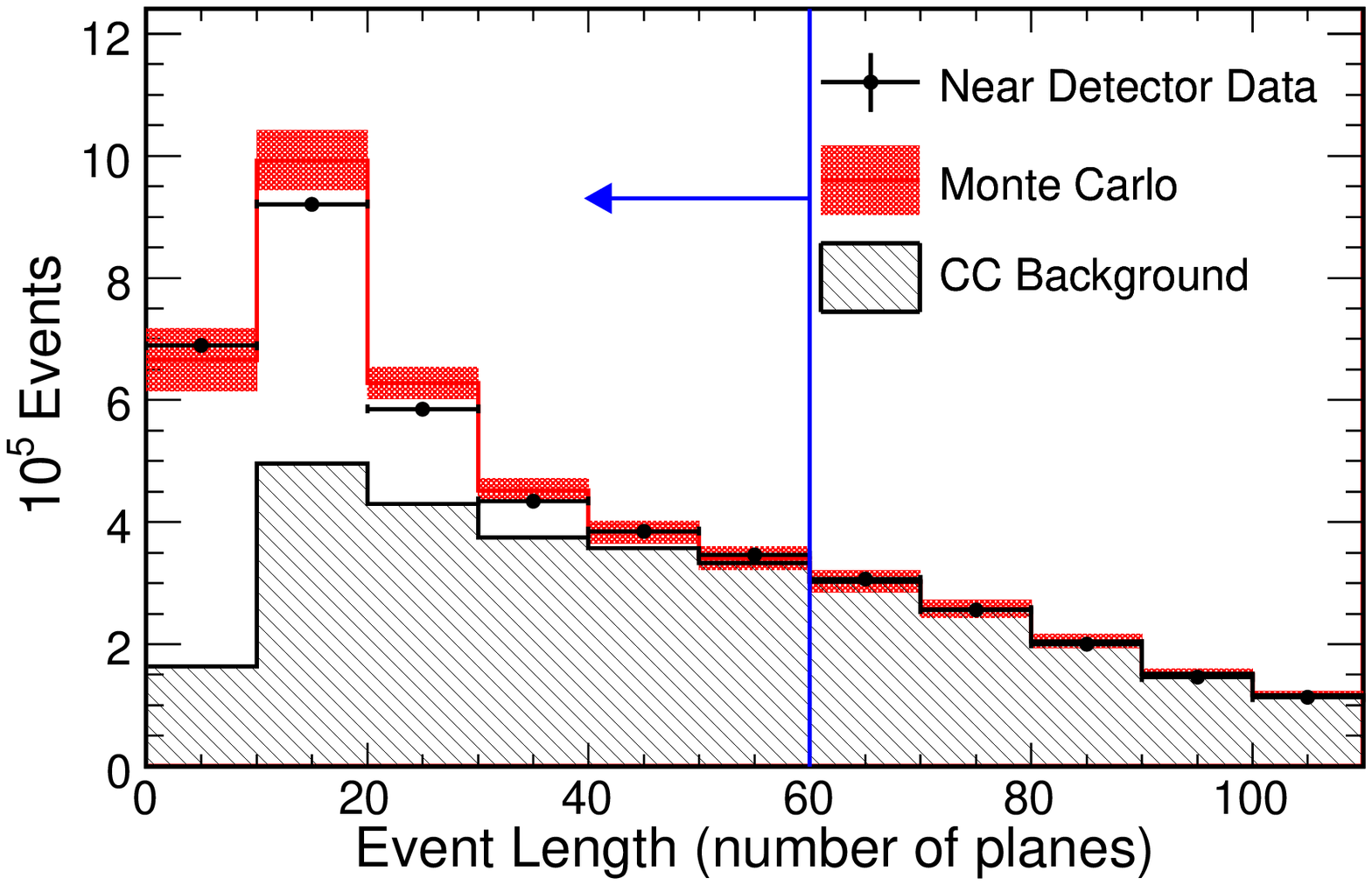}
        \put(12.8,54){\Large\bf a}
\end{overpic}
\begin{overpic}[width=1.05\linewidth]{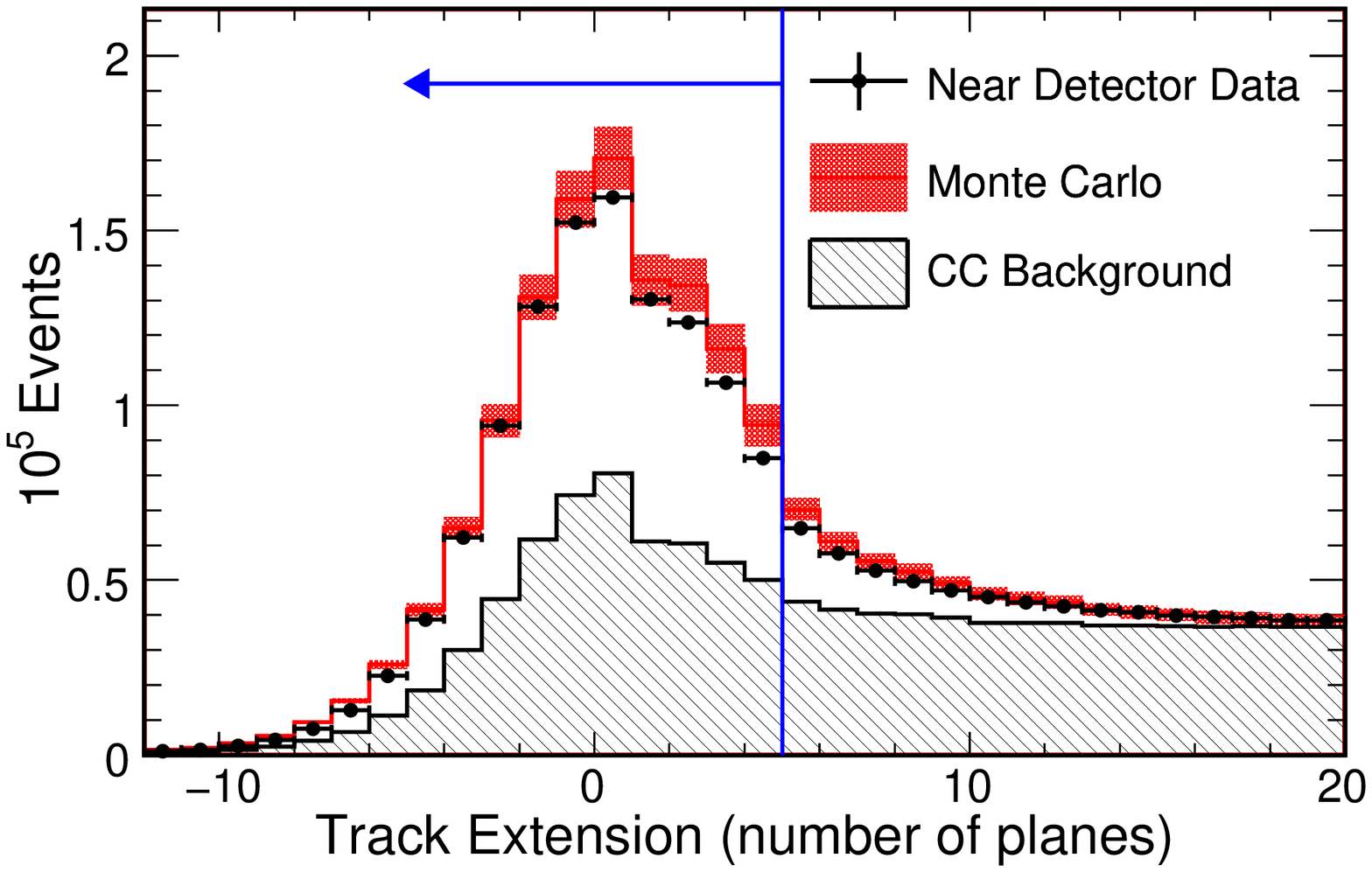}
        \put(12.8,52){\Large\bf b}
\end{overpic}
\caption{Comparisons of near-detector data with Monte Carlo predictions for distributions of the variables (a) {\it event length} and (b) {\it track extension}.  The data quality requirements described in Sec.~\ref{sec:Selection} are applied. Systematic uncertainties are displayed as shaded bands on the Monte Carlo expectation. Events selected as neutral-current-like are indicated by the arrows. }
\label{fig:NCPID}
\end{center}
\end{figure}

\begin{figure}
\begin{center}
\begin{overpic}[width=1.05\linewidth]{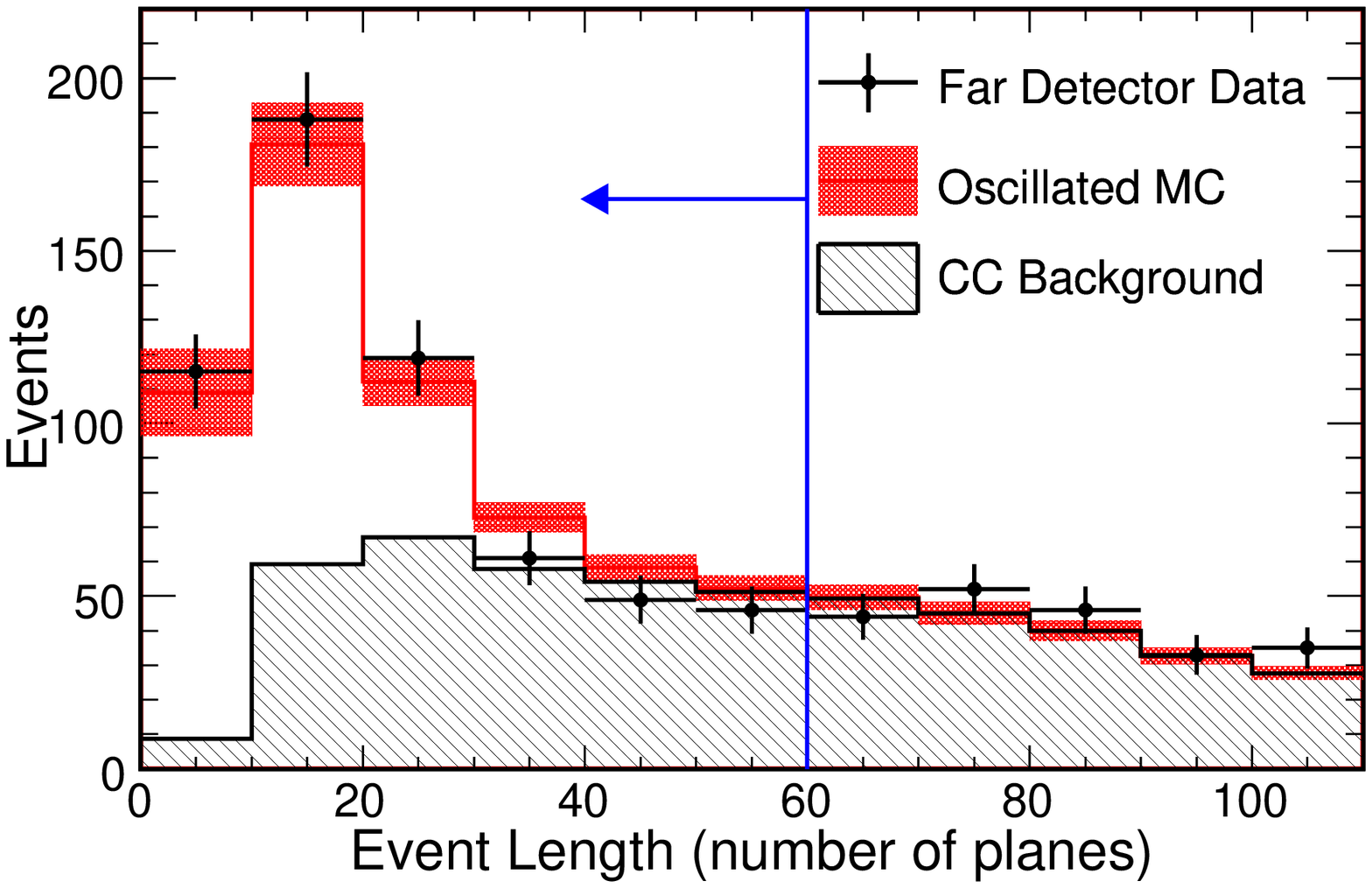}
        \put(12.8,54){\Large\bf a}
\end{overpic}
\begin{overpic}[width=1.05\linewidth]{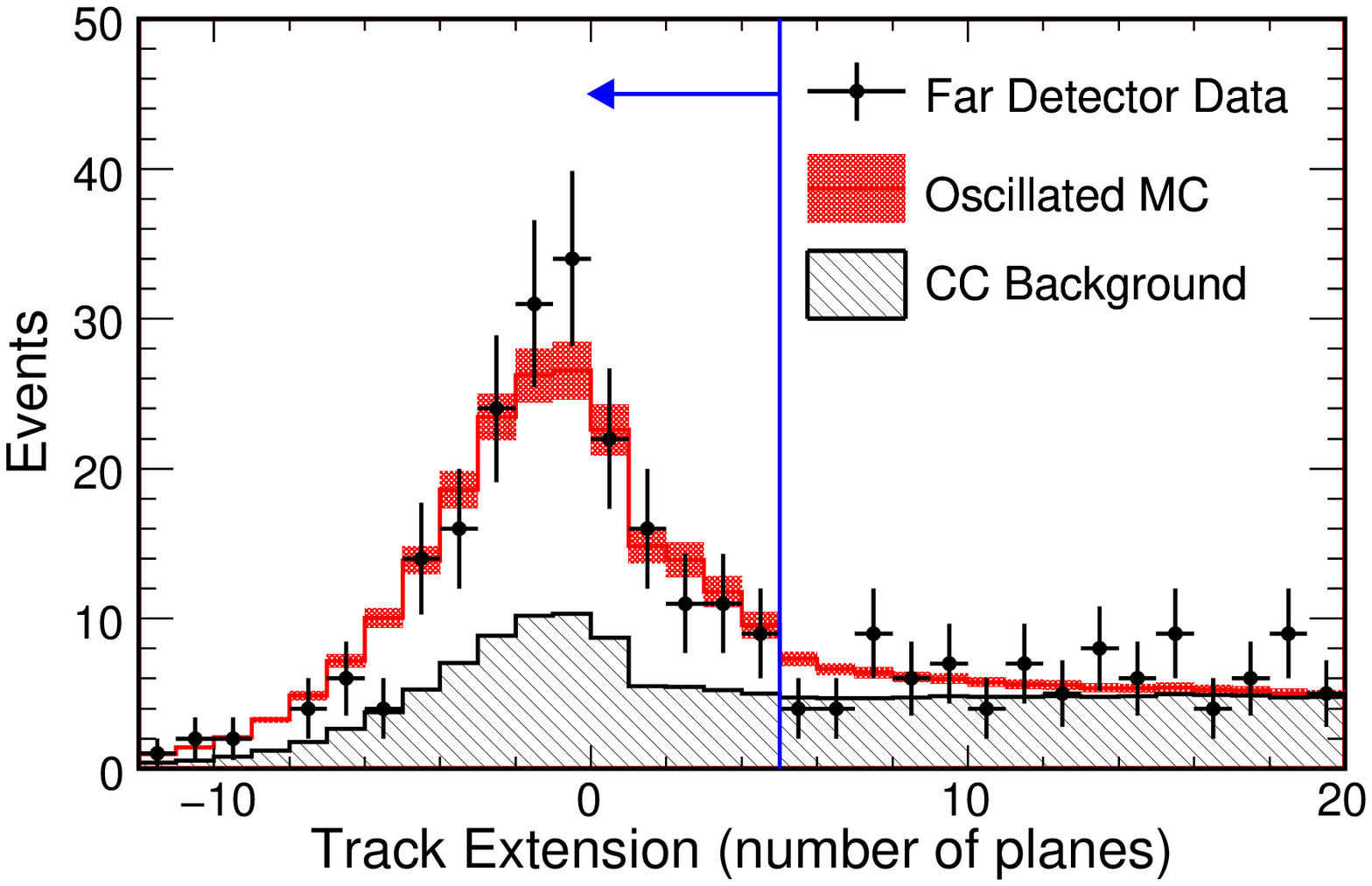}
        \put(12.8,52){\Large\bf b}
\end{overpic}
\caption{Far-detector data versus predictions from the Monte Carlo (MC) simulation including $\numu\rightarrow\nutau$ oscillations, for distributions of the classification variables (a) {\it event length} and (b) {\it track extension}.  The data quality requirements of Sec.~\ref{sec:Selection} are applied.  The shaded bands show the systematic errors on the MC predictions. The arrows indicate events selected as neutral-current-like.}
\label{fig:NCPIDfar}
\end{center}
\end{figure}

 Distributions for the event length and track extension classification variables for data of the near-detector are shown in Figs.~\ref{fig:NCPID}a and~\ref{fig:NCPID}b.   The data are plotted together with the prediction of the MINOS Monte Carlo simulation, which adequately reproduces the shapes of the classification-variable distributions.
Comparisons of distributions in the far-detector for the same event classification variables are shown in Figs.~\ref{fig:NCPIDfar}a and~\ref{fig:NCPIDfar}b. Here, the Monte Carlo simulation uses oscillation parameter values obtained from the most recent MINOS charged-current measurement, $|\Delta m^2_{32}|=\unit[2.43\times10^{-3}]{\rm{eV^2}}$ and $\sin^22\theta_{23}=1$~\cite{ref:CCPRL}.

\begin{figure}
\begin{center}
  \includegraphics[width=1.05\linewidth]{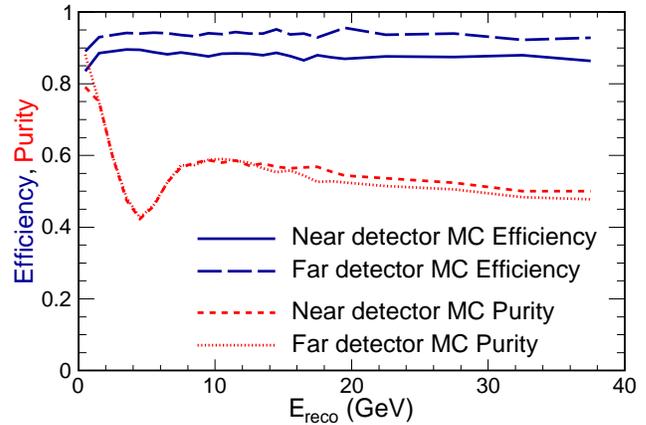}
  \caption{ Selection efficiency and sample purity for Monte Carlo (MC) events selected as neutral-current candidates in the near and far-detectors, as a function of reconstructed event energy. Detector selection, fiducial volume and neutral-current/charged-current separation requirements have been applied. The shapes of both efficiency and purity distributions are observed to be very similar for each of the two MINOS detectors.}
\label{fig:NCEffPur}
\end{center}
\end{figure}

\begin{figure}
\begin{center}
  \includegraphics[width=1.05\linewidth]{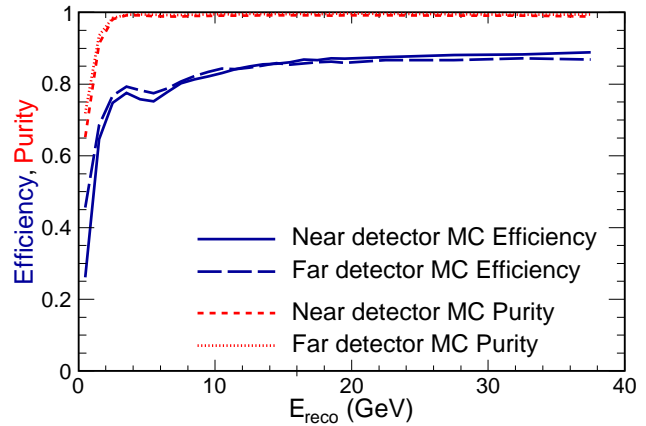}
  \caption{Distributions of selection efficiency (lower curves) and sample purity (upper curves) versus reconstructed event energy, for Monte Carlo (MC) events selected as charged-current candidate events in the near-detector and in the far-detector.}
\label{fig:CCEffPur}
\end{center}
\end{figure}

Efficiencies and purities for the neutral-current and charged-current event samples for both detectors are displayed in Figs.~\ref{fig:NCEffPur} and \ref{fig:CCEffPur}. The classified neutral-current samples have nearly identical purities for the near and far-detectors. The far-detector purity curve is computed for the case of null neutrino oscillations. The neutral-current sample efficiencies have identical trends but exhibit a constant relative offset over the full range of reconstructed event energies, $E_{\mathrm{reco}}$, due to the special near-detector selection criteria. The minima observed in sample purities for both detectors near the peak of the focussed neutrino spectrum, $E_\mathrm{reco}\approx\unit[4]{GeV}$, reflect the abundant rate and consequently increased background from \num~charged-current events.  

\begin{figure}
  \includegraphics[width=\linewidth]{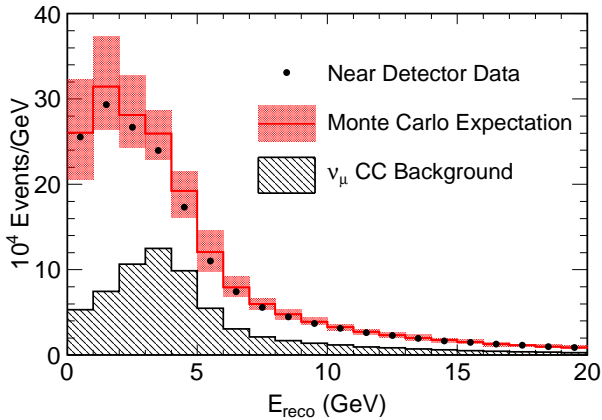}
  \caption{Distribution of reconstructed visible energy for selected neutral-current events in the near-detector, for the data (solid points) versus the Monte Carlo prediction (open histogram). The systematic errors (1$\sigma$) for the Monte Carlo are shown by the shaded band. Also shown is the Monte Carlo prediction for the background of misidentified charged-current events in the near-detector sample (hatched histogram).}
  \label{fig:nd_spectrum}
\end{figure}

\begin{figure}
  \includegraphics[width=\linewidth]{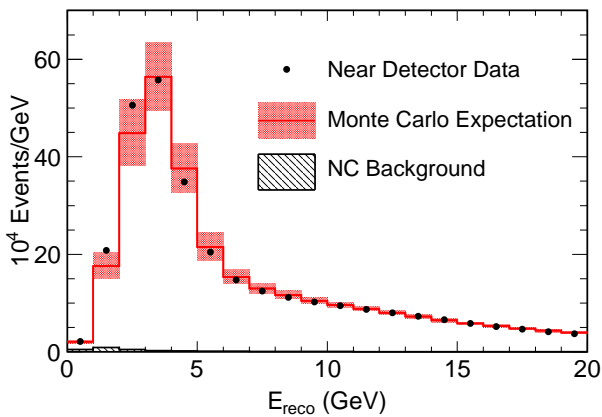}
  \caption{The reconstructed energy spectrum for selected charged-current events of the near-detector, for the data (solid points) versus the Monte Carlo (open histogram). Systematic errors at $1\sigma$ for the Monte Carlo are indicated by the shaded band; the hatched histogram (lower left) shows the Monte Carlo expectation for the small background contribution by misidentified neutral-current events to this sample.}
  \label{fig:nd_spectrum_CC}
\end{figure}

The resulting reconstructed energy spectra for the neutral-current and charged-current samples in the near-detector are shown in Figs.~\ref{fig:nd_spectrum} and \ref{fig:nd_spectrum_CC} respectively. The figures show that the selected neutral-current sample includes a sizable background contribution from misidentified charged-current events, in contrast to the selected charged-current sample which contains very little neutral-current background.   For both samples, the data points are seen to fall within or near the 1~standard deviation range of the systematic uncertainty of the Monte Carlo simulation.

\section{Far-Detector Prediction}\label{sec:Prediction}
The predictions of the energy spectra of the neutral-current and charged-current samples at the far-detector are based on the observed near-detector data and make use of the expected relationship between the neutrino fluxes at the two sites.  The process of making the predictions is called ``extrapolation''  and may be viewed as making corrections to the simulation of interactions in the far-detector based on the energy spectrum measured in the near-detector.  

This analysis uses an extrapolation technique called the ``far over near'' (F/N) method~\cite{ref:CCPRD,ref:KoskinenThesis}. This method makes the prediction of the far-detector spectrum by taking the product of two quantities.  The first quantity is the ratio of the expected number of events from the Monte Carlo simulation for each energy bin in the far-detector and near-detector spectra.  The expected number of events for each energy bin in the far-detector spectra depends on the composition of event types entering the samples and the corresponding oscillation probabilities for that energy.  The second quantity is the number of observed near-detector data events. The F/N method prediction is robust against distortions arising from differences between data and Monte Carlo simulation in the near-detector as these distortions are translated to the far-detector and do not affect the oscillation measurement~\cite{ref:CCPRD}. 
 
For example, the extrapolation for the neutral-current spectrum accounts for both the signal neutral-current events and the background charged-current events from each neutrino flavor.  The extrapolation also addresses the ways in which the backgrounds change between the near and far-detectors.  Specifically, for the case of the \num~charged-current component of the neutral-current and charged-current
samples, the F/N extrapolation predicts the number of events at the far-detector for the $i$-th bin of reconstructed energy to be
\begin{equation}
  \displaystyle
  F_{i}^{\rm{prediction}}=N_{i}^{\rm{data}}\left(~\frac{\displaystyle
  \sum_{x}\sum_{j}F_{ij}^{\rm{MC}}~P_{\nu_\mu\rightarrow\nu_{x}}(E_j)}{N_{i}^{\rm{MC}}}\right),
  \label{eq:FoverN}
\end{equation}
where $N_{i}^{\rm{data}}$ is the number of selected events in the $i$-th reconstructed energy bin in the near-detector and $N_{i}^{\rm{MC}}$ is the number of events expected in that bin from the near-detector Monte Carlo simulation. The $F_{ij}^{\rm{MC}}$ represents the number of events expected from the far-detector Monte Carlo simulation in the $i$-th bin of reconstructed energy and $j$-th bin of true neutrino energy. In the equation, $E_{j}$ is the true neutrino energy and $P_{\nu_\mu\rightarrow\nu_{x}}$ the probability of muon-neutrino transition to any other flavor.  

In particular, for the neutral-current spectrum, the extrapolation must take neutrino oscillations into account to properly characterize the predominant background arising from misidentified charged-current $\numu$, and it must include the small spectral distortion resulting from misidentified charged-current $\nu_\tau$ and $\nu_e$ events.  Thus, there are five separate classes of events that must be extrapolated to the far-detector: (i)~genuine neutral-current interactions, (ii)~\numu~charged-current interactions, (iii)~\nutau~charged-current interactions, (iv)~possible \nue~charged-current interactions originating from \numu~oscillations, and (v)~charged-current \nue~interactions initiated by the intrinsic \nue~beam component.  The muon neutrinos in the simulation include oscillations and are integrated in bins of reconstructed energy to account for the changing background.  Oscillations of the intrinsic beam $\nue$ into $\numu$ are not taken into account as those $\nue$ comprise only 1.3\% of the neutrinos in the beam and mixing angle for such oscillations is so small as to make the contributions from that oscillation mode negligible.

The five classes of simulated events are used to construct individual two-dimensional histograms of true neutrino energy versus reconstructed energy. In each of these histograms, all the events in an individual bin of true energy are multiplied by the same survival or transition probability. After applying oscillations the simulated events are converted to a reconstructed energy spectrum by integrating across all the true energy bins for each bin of reconstructed energy, producing the sum in Eq.~(\ref{eq:FoverN}). The reconstructed energy spectra for each separate data set, signal and backgrounds, are added together into one spectrum. 
%The number of simulated events in each bin of reconstructed energy is then divided by the number of simulated events in the corresponding energy bin in the near-detector to produce the F/N ratio. In the ratio, the number of events in the near-detector is several orders of magnitude larger than the number of events expected in the far-detector for the same exposure. The far-detector neutrino energy spectrum prediction is the bin-by-bin product of the F/N ratio and the near-detector data.
%Because neutrinos oscillate as a function of their true energy, there is a concern that applying the same survival probability to every event in a bin of true energy will introduce a systematic uncertainty to the prediction. The width of the true energy bins was chosen to be 0.12~GeV so that no measurable energy shift is introduced in approximating all the neutrinos in a single bin as having the same energy. 
%Along with the bin width, the use of a high-statistics Monte Carlo sample reduces uncertainty stemming from applying the survival probabilities in a bin-by-bin fashion.

\section{Systematic Uncertainties}\label{sec:Systematics}
The principal sources of systematic uncertainties in these analyses are: (i) {\it absolute scale of the hadronic energy}; (ii) {\it relative calibration of the hadronic energy} between the two detectors; (iii) {\it relative normalization} between the two detectors; (iv) {\it charged-current background} in selected neutral-current events; and (v) {\it uncertainties due to the near-detector selection requirements} in the near-detector event counts. Monte Carlo studies have been performed where, for each single uncertainty, the Monte Carlo spectrum is varied by $\pm$1 standard deviation independently, in order to estimate the effect of each on the extrapolated spectrum. Beam and cross section uncertainties that are common to the two detectors effectively cancel when using the F/N extrapolation.

The absolute hadronic energy scale has an uncertainty of 12\%.  This value is a combination of the uncertainty in the hadronization model and intranuclear effects (10\%) and uncertainty of the detector response to single hadrons (6\%).  An uncertainty of 3\% on the relative energy scale between the two detectors was determined from cross-calibration studies using stopping muons~\cite{ref:CCPRD}. The relative normalization between the two detectors has an uncertainty of 4\%. This is a combination of the uncertainties due to fiducial mass, live time, and reconstruction differences between the two detectors. 

To evaluate the uncertainties due to the near-detector selection, the requirement that the total number of reconstructed strips in an event is at least four was shifted by $\pm1$ strip. The effects on the reconstructed energy spectrum were determined for each shift. The uncertainty has been estimated to be 15.2\% for $E_{\text{reco}}<$\;0.5 \;GeV, 2.9\% for 0.5\;$<E_{\text{reco}}<$\;1\;GeV, 0.4\% for 
1\;$<E_{\text{reco}}<$\;1.5\;GeV and is negligible for higher visible energies.

 The uncertainty in the number of charged-current background events is determined using near-detector data taken in several different beam configurations, namely, (i) {\it horns-off}, in which there was no current in the magnetic horns; (ii) {\it medium energy}, in which the  target is moved upstream from the first horn by 100 cm; and (iii) {\it high energy}, in which the target is moved upstream from the first horn by 250 cm.  For each of these beam configurations, described in further detail in Ref.~\cite{ref:CCPRD}, the charged-current background component has a considerably different spectrum from the one obtained in the low-energy (LE) configuration. The charged-current background component in the neutral-current spectrum can thus be determined by using the observed differences in energy spectrum between the low-energy beam configuration and each of the other beam configurations along with information from Monte Carlo simulation of each configuration. In the low-energy configuration, the number of selected near-detector neutral-current events $N^\mathrm{LE}$ can be written as the sum of true neutral-current events and background charged-current events in that beam configuration: 

\begin{equation}
N^\mathrm{LE} = N_{\rm{nc}}^{\rm{LE}} + N_{\rm{cc}}^{\rm{{LE}}}
\label{eq:hornoff1}
\end{equation} 
For an alternate beam configuration ``Alt'', the number of selected neutral-current events may be 
written as
\begin{equation}
N^{\rm{Alt}}= r^{\rm{Alt}}_{\rm{nc}} \cdot N_{\rm{nc}}^{\rm{LE}} + r^{\rm{Alt}}_{\rm{cc}} \cdot N_{\rm{cc}}^{\rm{LE}},
\label{eq:hornoff2}
\end{equation} 
where 
$r^{\rm{Alt}}_{\rm{nc}}=\left. N_{\rm{nc}}^{\rm{Alt}} \middle/ N_{\rm{nc}}^{\rm{LE}} \right.$ and $r^{\rm{Alt}}_{\rm{cc}}=\left. N_{\rm{cc}}^{\rm{Alt}} \middle/ N_{\rm{cc}}^{\rm{LE}} \right.$ are determined from the Monte Carlo simulation. Equations~(\ref{eq:hornoff1}) and (\ref{eq:hornoff2}) can be solved to yield the solutions:
\begin{eqnarray}
N_{\rm{cc}}^{\rm{LE}}&=& \left.\left(N^{\rm{Alt}}-r^{\rm{Alt}}_{\rm{nc}}\cdot N^{\rm{LE}}\right)\middle/ \left(r^{Alt}_{\rm{cc}}-r^{\rm{Alt}}_{\rm{nc}}\right) \right.  \nonumber \\
N_{\rm{nc}}^{\rm{LE}}&=& \left.\left(N^{\rm{Alt}}-r^{\rm{Alt}}_{\rm{cc}}\cdot N^{\rm{LE}}\right)\middle/ \left(r^{Alt}_{\rm{nc}}-r^{\rm{Alt}}_{\rm{cc}}\right) \right.
\label{eq:hornoff3}
\end{eqnarray}
The final estimate on the $N_{\rm{cc}}^{\rm{LE}}$ background number results from the weighted average of the three different solutions of Eq.~(\ref{eq:hornoff3}) when the LE beam data is compared with each of the other beam configurations. 
The uncertainty on the $N_{\rm{cc}}^{\rm{LE}}$ background number is $12\%\pm2\%$. Therefore, the uncertainty in the number of charged-current background events is conservatively taken to be 15\% at all energies at the near and far-detectors.

\section{Search for Active Neutrino Disappearance}\label{sec:3Flavor}
\begin{figure}
  \includegraphics[width=1.05\linewidth]{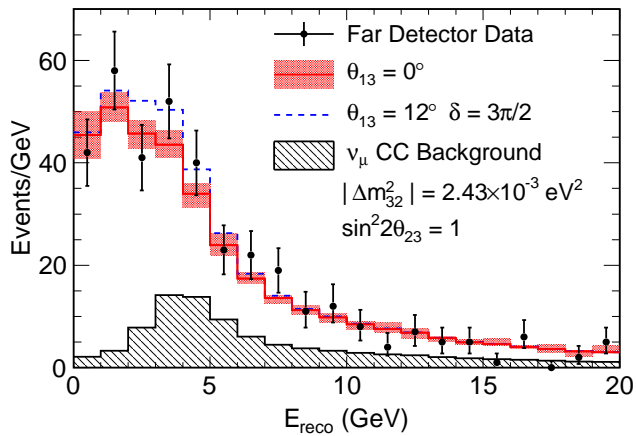}
  \caption{The reconstructed energy spectrum of neutral-current selected events at the far-detector (points with statistical uncertainties). The Monte Carlo prediction assuming standard three-flavor oscillations is also shown, both with (dashed line) and without (solid line) $\nu_e$ appearance at the CHOOZ limit. The shaded region indicates the 1 standard deviation systematic uncertainty on the prediction. The hatched region shows the Monte Carlo prediction for the background of misidentified charged-current events in this sample.}
  \label{fig:fd_spectrum}
\end{figure}

The data collected in the near and far-detectors have been classified as either neutral-current or charged-current events using the selections described in the previous sections.  A total of 388 data events are selected as neutral-current in the far-detector.  The measured and predicted $E_{\text{reco}}$ spectra at the far-detector are shown in Fig.~\ref{fig:fd_spectrum}.  The prediction assumes that oscillations occur only among the three active flavors and uses the values of $|\Delta m^{2}_{32}|$ and $\theta_{23}$ previously measured by MINOS~\cite{ref:CCPRL}.  

Although the present analysis is not capable of isolating an electron neutrino appearance signal, it must take $\numu\rightarrow\nue$ oscillations into account because the classification criteria of this analysis include \nue~charged-current interactions in the neutral-current enriched sample with nearly 100\% efficiency. This accounting is done by comparing the observed neutral-current spectrum to two predictions, one that assumes null \nue~appearance, and another that assumes an upper limit for the $\nue$ appearance rate in the far-detector calculated with the normal neutrino-mass hierarchy, $\theta_{13} = 12^\circ$ and $\delta = 3\pi/2$.  The choice of $\theta_{13}$ corresponds to the 90\% confidence level upper limit established by the CHOOZ reactor experiment~\cite{Apollonio:2002gd} for the $|\Delta m^{2}_{32}|$ value measured by MINOS~\cite{ref:CCPRL}; the choice of $\delta$ maximizes the $\nue$ appearance for the chosen value of $\theta_{13}$.  As seen in Fig.~\ref{fig:fd_spectrum}, the observed spectrum matches the prediction based on oscillations among the three active flavors very well over the full range of allowed values of $\theta_{13}$.

%The total number of neutral-current events selected in the far-detector is 388, whereas the expectation from standard three-flavor neutrino models is 377.2$\pm$19.4$\pm$18.5 (stat~$\pm$~syst)

The agreement between the observed and predicted neutral-current spectra is quantified using a statistic, $R$:  
\begin{equation}
R \equiv \frac{N_{\text{Data}}-B_{\text{CC}}}{S_{\text{NC}}},
\label{eq:fdef}
\end{equation}
where, within a given energy range, $N_{\text{Data}}$ is the observed event count, $B_{\text{CC}}$ is the extrapolated charged-current background from all flavors, and $S_{\text{NC}}$ is the extrapolated number of neutral-current interactions~\cite{ref:NCPRL}. The values of $S_{\text{NC}}$ and contributions to $B_{\text{CC}}$ are shown in Table~\ref{table:nums}.  

The disappearance of $\num$ occurs mainly for true neutrino energies $< 6$~GeV~\cite{ref:CCPRL}.  While the true energy of a neutrino interacting through the neutral-current process cannot be measured, the data can be separated into two samples whose reconstructed energy roughly discriminates between neutrinos with true energies greater than and less than 6~GeV. According to the Monte Carlo simulation, events with $E_{\text{reco}}<3$~GeV have a median true neutrino energy of 3.1~GeV while events with $3~\text{GeV}<E_{\text{reco}}<120$~GeV have a median true neutrino energy of 7.9~GeV.  The values of $R$ calculated for these ranges in $E_{\text{reco}}$ are shown in Table~\ref{table:nums}.  For all data samples, $R$ differs from unity by less than 1~standard deviation.  The effect of $\nue$ appearance on the value of $R$ is treated as an uncertainty in this analysis and is indicated by the last uncertainty in Table~\ref{table:nums}.  Because $\nue$-charged-current events are almost always identified as neutral-current, the effect of $\nue$~appearance at the far-detector is to decrease the value of $R$, since the predicted background will be larger than for the case of null $\nue$~appearance. 

The measured values of $R$ for each energy range indicate that neutrino oscillations among the active flavors describes the observed data quite well.  Over the full energy range, $0-120$~GeV, a value of $R=1.04\pm0.08\text{(stat.)}\pm0.07\text{(syst.)}-0.10(\nue)$ is measured, corresponding to a depletion of the total neutral-current event rate assuming null (maximally-allowed) \nue~appearance of less than $8\%~($18\%) at 90\% confidence level.  The following sections explore the extent to which the data allow oscillations between the active flavors and one sterile neutrino or oscillations in combination with neutrino decay.

\begin{table}
\begin{tabular}{llcccc}
\hline
$E_{\text{reco}}$ (GeV)      & $N_{\text{Data}}$ & ~~$S_{\text{NC}}$~~ & ~~$B^{\num}_{\text{CC}}$~~ & ~~$B^{\nut}_{\text{CC}}$~~ & ~~$B^{\nue}_{\text{CC}}$~~ \\ \hline
$0-3$       &141&125.1&13.3&1.4&2.3~~(12.4)   \\
$3-120$  &247&130.4&84.0&4.9&16.0 (32.8) \\
\hline 
$0-3$ &\multicolumn{5}{l}{$R=0.99\pm0.09\pm0.07-0.08$} \\
$3-120$ &\multicolumn{5}{l}{$R=1.09\pm0.12\pm0.10-0.13$} \\
$0-120$ &\multicolumn{5}{l}{$R=1.04\pm0.08\pm0.07-0.10$} \\
\hline 
\end{tabular}
\caption{\label{table:nums} Values of $R, N_{\text{Data}}, S_{\text{NC}},$ and the contributions to $B_{\text{CC}}$ for various reconstructed energy ranges.  The numbers in parentheses are calculated including $\nue$ appearance at the CHOOZ limit, as discussed in the text.  The first uncertainty in the value of $R$ shown is the statistical uncertainty, the second is systematic uncertainty, and the third is due to possible $\nue$ appearance.}  
\end{table}

\section{Oscillations Including a Sterile Neutrino Flavor}\label{sec:4Flavor}
Mixing of the three active neutrino flavors with one sterile neutrino requires the addition of one mass eigenstate.  The probability for mixing between any two flavors is described by the neutrino mixing matrix~\cite{pmns}, $U$, which defines the rotation from the mass basis into the flavor basis.   Thus, the mixing matrix needs to be expanded by one row and one column to accommodate the additional mass and flavor states.  The expanded $4\times4$ mixing matrix contains six mixing angles and six phases, with three of the phases being Majorana phases that are not relevant to an oscillation experiment~\cite{Schechter:1980gr}.  The matrix can be written as a  product of six independent rotation matrices about the Euler axes $R_{ij}$, where $ij$ refers to the plane in which a particular rotation takes place. The ordering of the rotation matrices is arbitrary, reflecting that the mixing matrix can be parameterized in many ways. The ordering described below was chosen to make the analysis more straightforward.

MINOS is designed to precisely measure $|\Delta m^{2}_{32}|$ but has no sensitivity to $\Delta m^{2}_{21}$.  Consequently, the mass states 1 and 2 are treated as degenerate in this analysis.  When two mass states are degenerate the rotation in the $ij$-plane is unphysical and the corresponding mixing angle, $\theta_{ij}$, vanishes from the oscillation probabilities~\cite{Donini:2007yf}.   Additionally, the matrix should be of the form such that the $U_{e3}$ component of the mixing matrix goes to zero when $\theta_{13}=0\,^{\circ}$ to distinguish an electron component in the third mass eigenstate from effects of sterile neutrinos.  For those reasons, the general form of the mixing matrix used by the current analysis is written as
\begin{widetext}
\begin{eqnarray}
\nonumber
U & = & R_{34}(\theta_{34})R_{24}(\theta_{24},\delta_{2})R_{14}(\theta_{14})R_{23}(\theta_{23})R_{13}(\theta_{13},\delta_{1})R_{12}(\theta_{12},\delta_{3}) \\
& = & R_{34}(\theta_{34})R_{24}(\theta_{24},\delta_{2})R_{14}(\theta_{14})R_{23}(\theta_{23})R_{13}(\theta_{13},\delta_{1}),
\label{eq:gen}
\end{eqnarray}
\end{widetext}
where the $\delta_{k}$ are the three CP-violating Dirac phases and the last equality reflects the assumption of degeneracy in mass states 1 and 2.
%The rotation matricies have elements of the form
%\begin{equation}
%R^{pq}_{ij}(\theta_{ij},\delta_{k})=
%	\begin{cases}
%		\cos\theta_{ij} & p = q = i \text{ or } p = q= j, \\
%		1 & p = q \neq i \text{ and } p = q \neq j, \\
%		\sin\theta_{ij}e^{-i\delta_{k}} & p = i \text{ and } q = j, \\
%		-\sin\theta_{ij}e^{i\delta_{k}} & p = j \text{ and } q = i, \\
%		0 & \text{otherwise},
%	\end{cases}
%\label{eq:rotelement}
%\end{equation}
%where $p$ indicates the row of $R_{ij}$ and $q$ indicates the column~\cite{Dighe:2007uf}.
Thus, the mixing matrix can be written as,
\begin{widetext}
\begin{equation}
U =\begin{bmatrix}
U_{e1} & U_{e2} & c_{14}s_{13}e^{-i\delta_{1}}& s_{14} \\
U_{\mu 1} & U_{\mu 2}& -s_{14}s_{13}e^{-i\delta_{1}}s_{24}e^{-i\delta_{2}}+c_{13}s_{23}c_{24} & c_{14}s_{24}e^{-i\delta_{2}} \\
U_{\tau 1} & U_{\tau 2} & -s_{14}c_{24}s_{34}s_{13}e^{-i\delta_{1}}-c_{13}s_{23}s_{34}s_{24}e^{i\delta_{2}}+c_{13}c_{23}c_{34} & c_{14}c_{24}s_{34} \\
U_{s1} & U_{s2} & -s_{14}c_{24}c_{34}s_{13}e^{-i\delta_{1}}-c_{13}s_{23}c_{34}s_{24}e^{i\delta_{2}}-c_{13}c_{23}s_{34} & c_{14}c_{24}c_{34} \\
\end{bmatrix}.
\label{eq:genmatrix}
\end{equation}
\end{widetext}
Here $c_{ij} = \cos\theta_{ij}$ and $s_{ij} = \sin\theta_{ij}$, and only the elements of the matrix that appear in the oscillation probabilities given below have been expressed in terms of the mixing angles and phases.

\subsection{Oscillation Probabilities}

The oscillation probabilities are derived following the example in Ref.~\cite{ref:PDG}.  The evolution of a neutrino with initial flavor state $\alpha$ is given by
\begin{equation}
\ket{\nu_{\alpha}} = \sum_{j}U^{*}_{\alpha j}e^{-im_{j}^{2}L/(2E)}\ket{\nu_{j},0},
\end{equation}
where the sum is over the mass states, $E$ is the neutrino energy, $L$ is the distance traveled by the neutrino and $m_{j}$ is the mass of state $j$.  As the NuMI beam is almost entirely $\num$ or $\overline{\nu}_{\mu}$, only the amplitudes and probabilities for $\num\rightarrow\nu_{x}$, where $x$ represents $e,\mu,\tau,$ or $s$, are described.  The amplitude for $\num\rightarrow\nu_{x}$ is given by
\begin{equation}
A = \sum_{j=1}^{4}U^*_{\mu j}U_{xj}e^{-im^{2}_{j}L/(2E)}.
\label{eq:amp1}
\end{equation}
In this equation $U_{\mu j}$ is the element of the mixing matrix describing the coupling between the muon flavor state and mass state $j$.  By squaring the amplitudes and using the unitarity of $U$, one obtains the oscillation probabilities for the different channels,
\begin{eqnarray}
P(\nu_{\alpha}\rightarrow \nu_{\beta}) & = & \delta_{\alpha\beta} -4\sum_{j>i}\mathcal{R}(U^{*}_{\alpha j}U_{\beta j}U_{\alpha i}U^{*}_{\beta i})\sin^{2}\Delta_{ji} \nonumber \\ & & + 2\sum_{j>i}\mathcal{I}(U^{*}_{\alpha j}U_{\beta j}U_{\alpha i}U^{*}_{\beta i})\sin 2\Delta_{ji},
\label{eq:genprobsmall}
\end{eqnarray}
where $\Delta_{ji} \equiv (m^{2}_{j}-m^{2}_{i})L/(4E)$ and $\mathcal{R()}$, $\mathcal{I()}$ designate the real and imaginary parts of the amplitude products.  Because $\Delta_{21} \ll \Delta_{31}$ the first and second mass states  are treated as degenerate and the factors of $\sin\Delta_{21}$ can be set to 0.  This degeneracy also implies that $\Delta_{42} = \Delta_{41}$ and $\Delta_{32} = \Delta_{31}$.  Equation~(\ref{eq:genprobsmall}) can be expanded for the different oscillation scenarios,
\begin{widetext}
\begin{eqnarray}
\nonumber
P_{\nu_\mu\rightarrow\nu_\mu} & = & 1 - 4\biggl\{\Uo{\mu 3}(1-\Uo{\mu 3} \nonumber -\Uo{\mu 4})\sin^{2}\Delta_{31} + \Uo{\mu 4}\Uo{\mu 3}\sin^{2}\Delta_{43} + \Uo{\mu 4}(1-\Uo{\mu 3}-\Uo{\mu 4})\sin^{2}\Delta_{41}\biggr\}, \label{eq:oscprobs} \\ \nonumber
P_{\nu_\mu\rightarrow\nu_\alpha} & = & 4\mathcal{R}\biggl\{\Ut{\mu 3}{\alpha 3}\sin^{2}\Delta_{31} + \Ut{\mu 4}{\alpha 4}\sin^{2}\Delta_{41} + U^{*}_{\mu 4}U_{\alpha 4}U_{\mu 3}U^{*}_{\alpha 3}(\sin^{2}\Delta_{31} -\sin^{2}\Delta_{43}+\sin^{2}\Delta_{41}) \biggr\} \\
& & + 2\mathcal{I}\biggl\{U^{*}_{\mu 4}U_{\alpha 4}U_{\mu 3}U^{*}_{\alpha 3}(\sin2\Delta_{31} - \sin2\Delta_{41} + \sin2\Delta_{43})\biggr\},\\ \nonumber
\end{eqnarray}
\end{widetext}
where $\alpha = e, \tau, \text{ or } s$ and the orthogonality constraint $\sum_{i}U_{ai}U^{*}_{bi} = 0$ has been used to eliminate the matrix elements corresponding to the first and second mass states.

As can be seen from Eqs.~(\ref{eq:genmatrix})~and~(\ref{eq:oscprobs}), mixing between three active and one sterile neutrino in the most general case involves ten parameters which are mostly unknown: five mixing angles, three mass-squared splittings, and two {\it CP}-violating phases.  The following simplifying assumptions are made to reduce the number of possible parameters. First, $\theta_{13}$ is eliminated as a free parameter by only considering two values,  $\theta_{13} = 0^{\circ}$ or $\theta_{13} = 12^{\circ}$, where the latter is the CHOOZ limit at the MINOS measured value of $|\Delta m^{2}_{32}|$. When $\theta_{13}$ is set to the CHOOZ limit, $\delta_{1}$ is taken to be $3\pi/2$ as that is the value of the {\it CP}-violating phase that maximizes the $\nue$ appearance probability.  Second, the additional {\it CP}-violating phase is removed by setting $\delta_{2} = 0$ as MINOS has no sensitivity to this phase.  Finally, nonzero values of $\theta_{14}$ do not measurably change the oscillation probabilities, so that angle is set to $0^\circ$ as well. 

\begin{figure}
  \includegraphics[width=1.05\linewidth]{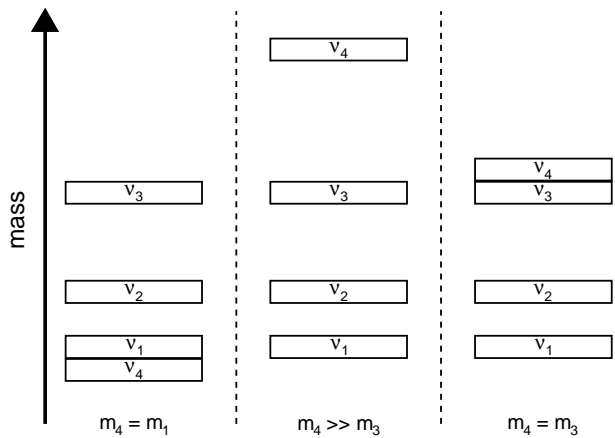}
  \caption{Schematic representation of the possible mass spectra for models including four mass eigenstates.  In the left panel, the first and fourth eigenstates are degenerate, the center panel has the fourth mass eigenstate much heavier than the third, and the right panel has the third and fourth eigenstates as degenerate.  Only the scenarios illustrated in the left and center panels are considered in this analysis.}
  \label{fig:massspect}
\end{figure}

In addition to the parameters represented in the mixing matrix and oscillation probability equations, there is one more free parameter, namely the mass of $\nu_{4}$ relative to the other mass states.  To limit the number of free parameters in the models, this analysis only examines possibilities for the relative mass of $\nu_{4}$ that allow the oscillation probabilities to depend only on $|\Delta m^{2}_{31}|$.  The possible hierarchies are illustrated in Fig.~\ref{fig:massspect} and only the normal mass hierarchy is shown.  An inverted mass hierarchy would change the relative locations of the pair $\nu_{1}$ and $\nu_{2}$ to $\nu_{3}$.  The fourth mass state could either be degenerate with the first mass state, as seen in the left panel of the figure, much more massive than the third mass state, as seen in the middle panel, or degenerate with the third mass state, as shown in the right panel.
In the scenario with the third and fourth eigenstates being nearly degenerate there would be no discernible mixing between the active and sterile components.  This consequence comes from the SNO results~\cite{Ahmad:2002jz} which indicate that any coupling between $\nu_{s}$ and the active neutrinos occurs in the third and fourth mass eigenstates. Therefore, in this  case the mass difference in which oscillations between active and sterile neutrinos could arise is too small to be observed in terrestrial baselines and it is not explored further in this analysis.  

An alteration of the oscillation probability of Eq.~(\ref{eq:oscprobs}) is predicted for neutrinos which traverse dense matter.  By virtue of their neutral-current coherent forward scattering with nucleons, active neutrinos acquire an effective matter potential whereas sterile neutrinos do not. The effective potentials are identical for the $\nu_\mu$ and $\nu_\tau$ flavors so that  matter effects vanish for $\nu_\mu\rightarrow\nu_\tau$ mixing. For $\nu_\mu\rightarrow \nu_s$ oscillations, however, the overall nonzero matter potential yields MSW-type modifications to the mixing angle and oscillation length such that, in normal (inverted) hierarchy scenarios, the oscillation probability will be suppressed (enhanced) relative to $\nu_\mu\rightarrow\nu_\tau$~\cite{ref:MSW}.  Observations of neutrinos with energies above \unit[12]{GeV} that have traveled through several thousand kilometers of dense material, such as those described in Ref.~\cite{superktau1}, would be sensitive to this matter effect. By contrast, the beam-induced neutrinos monitored at the MINOS far site originate from a $E_\nu$~spectrum highly peaked between 2 and \unit[6]{GeV} and travel \unit[735]{km} through the Earth's crust.  For the effective mixing angles and oscillation lengths of neutrinos in the NuMI beam, the $\nu_\mu\rightarrow\nu_s$ matter effect only gives rise to subpercent distortions of the neutrino oscillation probability that are mostly confined to neutrino energies below \unit[2]{GeV}.  Distortions of this magnitude are of no consequence for the studies reported here, and so matter effects involving sterile neutrinos have been neglected.

\subsubsection{Active-sterile mixing when $m_{4} = m_{1}$}

In the $m_{4} = m_{1}$ case, the first and fourth mass eigenstates are assumed to be degenerate.  Because the first and second eigenstates are also treated as degenerate, the second and fourth states are degenerate as well.  These degeneracies allow one to set $\theta_{14}=\theta_{24}=0^{\circ}$ in Eq.~(\ref{eq:genmatrix}), which reduces the number of parameters in the model to four. There are no $\nue$ or $\num$ components in the fourth mass eigenstate, however there is a $\nus$ component in the third mass eigenstate.  Using these simplifications, the oscillation probabilities become
\begin{eqnarray}
\nonumber
P_{\nu_\mu\rightarrow\nu_\mu} & = &1 - 4\Uo{\mu 3}\biggl(1-\Uo{\mu 3}\biggr)\sin^{2}\Delta_{31},  \label{eq:oscprobs41} \\
P_{\nu_\mu\rightarrow\nu_{\alpha}}  & = & 4\Ut{\mu 3}{\alpha 3}\sin^{2}\Delta_{31}\biggr.   \\ \nonumber
\end{eqnarray}
This model is equivalent to the phenomenological model presented by MINOS in Ref.~\cite{ref:NCPRL}.

\subsubsection{Active-sterile mixing when $m_{4} \gg m_{3}$}

In the  $m_{4} \gg m_{3}$ case, the fourth mass eigenstate is assumed to be much larger than the third; consequently the values of $\sin^{2}\Delta _{41}$ and $\sin^{2}\Delta_{43}$ average to $\frac{1}{2}$.  Additionally, $\sin 2\Delta_{41}$ and $\sin 2\Delta _{43}$ average to 0.  In this model $\Delta m^{2}_{43}$ is assumed to be $\mathcal{O}$(eV$^{2}$) such that the regime of rapid oscillations and thus the averages mentioned are valid at the far-detector site, while ensuring no detectable depletion of $\num$ occurs at the near-detector.  Such models have recently received attention in the literature~\cite{Donini:2007yf,steriletheory2}.  Using the above simplifications reduces the number of parameters in this model by two, and allows Eq.~(\ref{eq:oscprobs}) to be written as
\begin{eqnarray}
\nonumber
P_{\nu_\mu\rightarrow\nu_\mu} & = &1 - 4\biggl\{\Uo{\mu 3}\biggl(1-\Uo{\mu 3}-\Uo{\mu 4}\biggr)\sin^{2}\Delta_{31} \nonumber \\
& & + \frac{\Uo{\mu 4}}{2}(1-\Uo{\mu 4})\biggr\}, \label{eq:oscprobs43} \nonumber \\
P_{\nu_\mu\rightarrow\nu_\alpha}  & = & 4\mathcal{R}\biggl\{\biggl(\Ut{\mu 3}{\alpha 3}+U^{*}_{\mu 4}U_{\alpha 4}U_{\mu 3}U^{*}_{\alpha 3}\biggr)\sin^{2}\Delta_{31} \nonumber \\
& &  + \frac{\Ut{\mu 4}{\alpha 4}}{2} \biggr\},
\end{eqnarray}
where the second term of Eq.~(\ref{eq:oscprobs}) does not appear because of the assumptions that $\theta_{14} = 0^{\circ}$ and $\delta_{2} = 0$.

\subsection{Fitting active-sterile oscillations to the data}
\begin{figure}
  \includegraphics[width=1.05\linewidth]{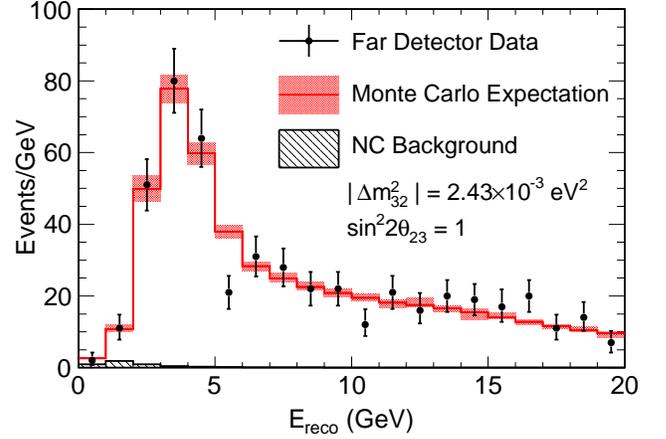}
  \caption{The reconstructed energy spectrum of charged-current selected events at the far-detector (points with statistical uncertainties). The Monte Carlo prediction assuming standard three-flavor oscillations (solid line) is shown with the 1 standard deviation systematic uncertainty on the prediction indicated by the shaded region. The prediction for the small background of misidentified neutral-current events in this sample (hatched region) is also shown.}
  \label{fig:fd_spectrum_CC}
\end{figure}

The data are compared to Monte Carlo predictions based on the probabilities in Eqs.~(\ref{eq:oscprobs41})~and~(\ref{eq:oscprobs43}) using the $\chi^{2}$ statistic appropriate to small sample sizes,
\begin{equation}
\chi^{2} = 2\sum^{N}_{i=1}\biggl[e_{i} - o_{i} + o_{i}\ln\frac{o_{i}}{e_{i}}\biggr] + \sum^{5}_{j = 1} \frac{\epsilon_{j}^2}{\sigma_{j}^2}.
\label{eq:chi2}
\end{equation}
Here $e_{i}$ is the expected number of events, assuming oscillations among four flavors,  in bin $i$ of the energy spectrum, and $o_{i}$ is the observed number of events in that bin.  The second sum is the contribution to $\chi^{2}$ from the parameters describing the systematic uncertainties; the nuisance parameter $\epsilon_{j}$ is the shift from the nominal fit value for the $j${\it -th} source of systematic uncertainty and $\sigma_{j}$ is the uncertainty associated with that source.  Both the neutral-current spectrum shown in Fig.~\ref{fig:fd_spectrum} and the charged-current spectrum shown in Fig.~\ref{fig:fd_spectrum_CC} are used to obtain the oscillation parameters that best fit the data.  The neutral-current spectrum provides information on the mixing angles for mixing between active and sterile neutrinos while the charged-current spectrum provides constraints on the mixing angle $\theta_{23}$ and the mass splitting $|\Delta m^{2}_{31}|$.  The charged-current-like spectrum of Fig.~\ref{fig:fd_spectrum_CC} is statistically consistent with that presented in Ref.~\cite{ref:CCPRL} given the different fiducial volumes and event separation procedures used in the two analyses.
The five systematic uncertainties described in Sec.~\ref{sec:Systematics} are included as nuisance parameters in the far-detector fits. By fitting for the systematic parameters simultaneously using both near and far-detector data, the effect of the uncertainties is substantially reduced due to significant cancellations of uncertainties between the two detectors.

The best-fit values for the mixing angles in the two models as well as the $\chi^{2}$ for each are shown in Table~\ref{tab:fit_pts}.
\begin{table}[!b]
  \begin{tabular}{|c*{6}{|c}|}
\hline
    Model & $\theta_{13}$   \bigstrut & $\chi^2$/D.O.F. & $\theta_{23}$  & $\theta_{24}$ & $\theta_{34}$ & $f_{s}$\\
% Model 12
    \hline
    \multirow{2}{*}{$m_4 = m_1$} & $0$   \bigstrut & 47.5/39 & $45.0^{+9.0}_{-8.9}$ &   & $0.1^{+28.7}_{-0.1}$  & 0.51\\
                                     & $12$ \bigstrut & 46.2/39 & $47.1^{+8.8}_{-11.0}$ &   & $23.0^{+22.6}_{-24.1}$ & 0.55\\
% Model 11
    \hline
    \multirow{2}{*}{$m_4\gg m_3$} & $0$   \bigstrut & 47.5/38 & ${45.0}^{+9.0}_{-8.9}$ & ${0.0}^{+7.2}_{-0.0}$ & ${0.1}^{+28.7}_{-0.1}$ & 0.52\\
                                 & $12$ \bigstrut & 46.2/38 & ${47.1}^{+8.8}_{-11.0}$ & ${0.0}^{+7.2}_{-0.0}$ & ${23.0}^{+22.6}_{-24.1}$ & 0.55\\
    \hline
  \end{tabular}
  \caption{Best-fit points and uncertainty ranges obtained for the active-sterile oscillation models. Results are shown with and without \nue~appearance at the CHOOZ limit.  All angles are given in degrees.  The quantity $f_{s}$ is defined as the fraction of disappearing $\nu_{\mu}$ that could transition to $\nu_{s}$ and is given at the 90\% C.L. in this table.  The values of $f_{s}$ in the $m_{4} \gg m_{3}$ model are evaluated for $E_{\nu} = \unit[1.4]{GeV}$.}
  \label{tab:fit_pts}
\end{table}

% CONTENTS OF OUTPUT FILES:
%
% DELTA43 MODEL (THETA13 = 0)
%
% Best fit  - chisq = 47.4723
%   theta34 = 0.00104615
%   theta23 = 0.785242
%   dmsq = 2.4414
%   theta24 = -1.79978e-17
%   norm = 0.02232
%   rel-hadr = 3.6e-05
%   abs-hadr = 0.021604
%   cc-bg = -0.0351
%   nd-clean = 0.2464
% Asymmetric errors in theta34 -0.0209021 0.50165
% Asymmetric errors in theta23 -0.155981 0.157864
% Asymmetric errors in theta24 -0.020923 0.125538
%
% DELTA43 MODEL (THETA13 = 0.21)
%
% Best fit  - chisq = 46.1816
%   theta34 = 0.40153
%   theta23 = 0.821334
%   dmsq = 2.4562
%   theta24 = -1.79978e-17
%   norm = 0.016704
%   rel-hadr = -0.002928
%   abs-hadr = 0.03388
%   cc-bg = -0.00612
%   nd-clean = 0.104
% Asymmetric errors in theta34 -0.421386 0.393795
% Asymmetric errors in theta23 -0.192073 0.153156
% Asymmetric errors in theta24 -0.020923 0.125538
%
% DELTA41 MODEL (THETA13 = 0)
%
% Best fit  - chisq = 47.4723
%   theta34 = 0.00104615
%   theta23 = 0.785242
%   dmsq = 2.4414
%   norm = 0.02232
%   rel-hadr = 3.6e-05
%   abs-hadr = 0.021604
%   cc-bg = -0.0351
%   nd-clean = 0.2464
% Asymmetric errors in theta34 -0.0209021 0.50165
% Asymmetric errors in theta23 -0.155981 0.157864
%
% DELTA41 MODEL (THETA13 = 0.21)
%
% Best fit  - chisq = 46.1816
%   theta34 = 0.40153
%   theta23 = 0.821334
%   dmsq = 2.4562
%   norm = 0.016704
%   rel-hadr = -0.002928
%   abs-hadr = 0.03388
%   cc-bg = -0.00612
%   nd-clean = 0.104
% Asymmetric errors in theta34 -0.421386 0.393795
% Asymmetric errors in theta23 -0.192073 0.153156
A list of the systematic effects for the fit parameters is presented in Table \ref{tab:systTable}.  The latter table shows that for each mixing angle evaluated in the four-neutrino models considered,  the uncertainties introduced by the five most significant sources of systematic error are relatively small compared to the total uncertainty ranges obtained from the fits, as summarized in Table~\ref{tab:fit_pts}. 
\begin{table*}
\begin{tabular}{|c|c|c|c|c|c|c|c|}
\hline
%Do the tricky table header
\multirow{2}{*}{Model} \bigstrut 
& \multirow{2}{*}{Parameter} \bigstrut 
& \multicolumn{6}{|c|}{Shift due to Systematic Uncertainty} \bigstrut \\
\cline{3-8} 
& & Absolute $\rm{E_{Had.}}$ \bigstrut 
& Relative $\rm{E_{Had.}}$\bigstrut 
& Normalization \bigstrut
& CC Background \bigstrut
& ND Selection \bigstrut 
& Total \\
%multirow requires empty elements for extra rows 
\hline
\multirow{2}{*}{$m_4 = m_1$} & $\Delta \theta_{23}$   \bigstrut & $0.3$& $0.6$ & $0.3$ & $0.1$  & $0.1$ & $0.7$ \\
                                     & $\Delta \theta_{34}$  \bigstrut & $3.6$ & $9.9$ & $12.6$ & $9.9$ & $9.9$ & $21.6$ \\
\hline
\multirow{3}{*}{$m_4 \gg m_3$} & $\Delta \theta_{23}$   \bigstrut & $0.2$ & $0.6$ & $0.1$  & $0.2$ & $0.2$ &  $0.7$\\
                                     & $\Delta \theta_{24}$ \bigstrut & $1.5$ & $2.1$ & $5.1$ & $0.3$ & $0.3$ & $5.7$\\
                          & $\Delta \theta_{34}$  \bigstrut & $4.5$ & $9.9$ & $6.3$ & $9.9$ & $9.9$ & $18.8$ \\
\hline
\multirow{2}{*}{Oscillations} & $\Delta \alpha$ (GeV/km)  \bigstrut & $2.54\times10^{-4}$ &  $0.70\times10^{-4}$ &  $6.25\times10^{-4}$ &  $1.23\times10^{-4}$  &  $1.15\times10^{-4}$ &  $6.99\times10^{-4}$ \\
                           {with decay}    & $\Delta \theta$  \bigstrut & $2.6$ & $3.7$ & $0.9$ & $4.0$ & $3.9$ & $7.2$\\
\hline
\end{tabular}
\caption{Summary of mixing-angle deviations introduced by the major systematic uncertainties from best-fit results in which systematic shifts have been neglected. Angular deviations, shown in degrees, are displayed for each mixing angle fitted, for each of the neutrino models analyzed in this work.}
\label{tab:systTable}
\end{table*}

\begin{figure}
\includegraphics[width=1.05\linewidth]{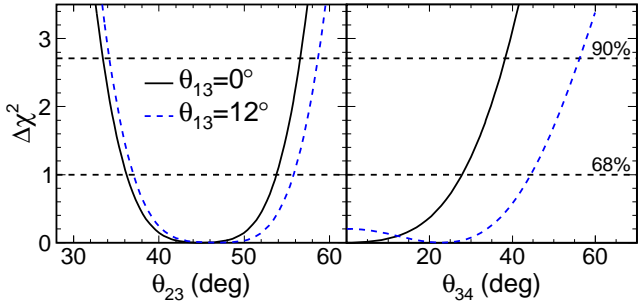}
\caption{Projections of $\Delta\chi^2$ as a function of the mixing angles for the $m_{4} = m_{1}$ model. The solid line contours are obtained with null $\nu_e$ appearance, whereas the dashed line contours include $\nu_e$ appearance at the CHOOZ limit. The ranges of values allowed at 68\% and 90\% confidence levels lie within contours below the horizontal dashed lines.}
\label{fig:delta41_projs}
\end{figure}

\begin{figure}
\includegraphics[width=1.05\linewidth]{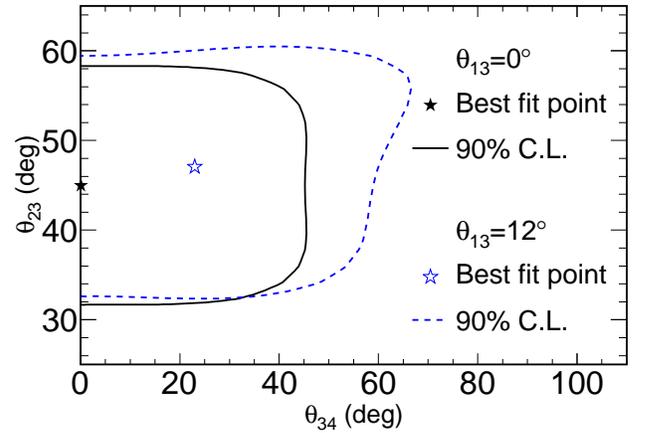}
\caption{Contours representing 90\% confidence level for the $m_{4} = m_{1}$ model. The solid line and best-fit point (solid symbol) are obtained assuming null $\nu_e$ appearance, whereas the dashed line and corresponding best-fit point (open symbol) are obtained with $\nu_e$ appearance set at the CHOOZ limit.}
\label{fig:delta41_contour}
\end{figure}

\begin{figure*}
\includegraphics[width=0.9\linewidth]{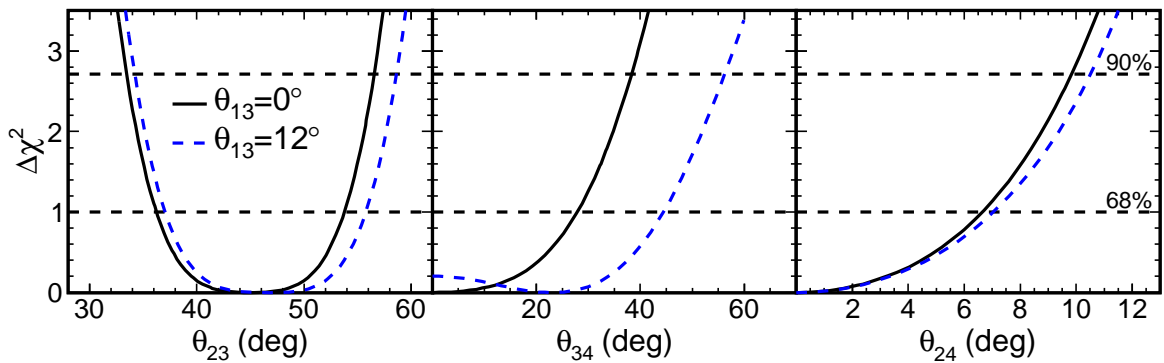}
\caption{Projections of $\Delta\chi^2$ as a function of the mixing angles for the $m_{4}\gg m_{3}$ model. The solid line is obtained for the case of null $\nu_e$ appearance whereas the dashed line represents solutions with $\nu_e$ appearance at the CHOOZ limit.  The ranges of values allowed at 68\% and 90\% confidence levels lie within contours below the horizontal dashed lines.}
\label{fig:delta43_projs}
\end{figure*}

\begin{figure}
\includegraphics[width=1.05\linewidth]{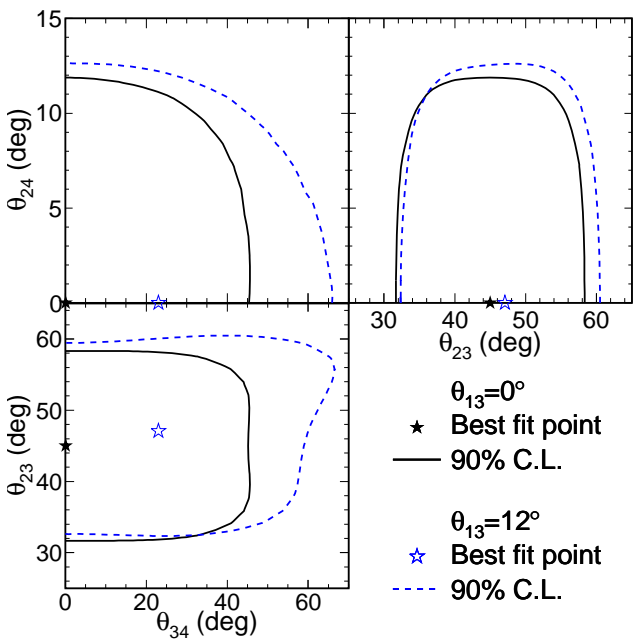}
\caption{Contours representing 90\% confidence level for the $m_{4}\gg m_{3}$ model. The solid line and best-fit point (solid symbol) are obtained for the case of null $\nu_e$ appearance, whereas the dashed line and corresponding best-fit point (open symbol) is obtained with $\nu_e$ appearance included with $\theta_{13}$ at the CHOOZ limit.}
\label{fig:delta43_contours}
\end{figure}

The best-fit values obtained for each model for the mass splitting, $|\Delta m^{2}_{31}|$, agree well with the result found in Ref.~\cite{ref:CCPRL}.  The one-dimensional projections of the $\Delta\chi^{2}$ between the best-fit point and the remaining points in the space are shown in Fig.~\ref{fig:delta41_projs} for the $m_{4} = m_{1}$ model. The two-dimensional 90\% confidence level contours for that model are shown in Fig.~\ref{fig:delta41_contour}. The projections and contours for the $m_{4} \gg m_{3}$ model are shown in Figs.~\ref{fig:delta43_projs}~and~\ref{fig:delta43_contours}.

As seen in these figures, $\theta_{34} < 38^\circ\,(56^\circ)$ at 90\% confidence level for the $m_{4} = m_{1}$ model. The number in parentheses represents the 90\% confidence level limit obtained when maximal $\nue$ appearance is allowed. For the $m_{4}\gg m_{3}$ model, $\theta_{24} < 10^\circ\,(11^\circ)$ and $\theta_{34} < 38^\circ\,(56^\circ)$ at the 90\% confidence level.  These limits indicate that any coupling between the active neutrinos and a sterile neutrino is submaximal.  Furthermore, the $\chi^{2}$ values indicate that these four-flavor models fit the data no better than oscillations among only the active neutrinos.

A straightforward method to quantify the coupling between the active and sterile neutrinos is to determine the fraction of disappearing $\num$ that transition to $\nus$.  That fraction is expressed as
\begin{equation}
 f_{s}\equiv\frac{P_{\nu_\mu\rightarrow\nu_s}}{1-P_{\nu_\mu\rightarrow\nu_\mu}}.
\label{eq:fsdef}
\end{equation}
For the $m_{4} = m_{1}$ model, the disappearance fraction $f_{s}$ is energy independent, as can be seen upon inserting the expressions from Eq.~(\ref{eq:oscprobs41})~into~Eq.~(\ref{eq:fsdef}). The 90\% confidence level limit for $f_{s}$ is determined by selecting a large number of test values of $\theta_{23}$ and $\theta_{34}$ from Gaussian distributions with mean and $\sigma$ given in Table~\ref{tab:fit_pts}.  The value of $f_{s}$ that is larger than 90\% of the test cases represents the limit.  The value corresponding to the 90\% confidence level is  $f_{s} < 0.51\,(0.55)$, with the value in parentheses indicating the value obtained for maximally-allowed $\nue$ appearance in the beam. This new limit on the value of $f_{s}$ represents a reduction of 33\% compared to the previous MINOS result without $\nue$ appearance~\cite{ref:NCPRL}. 
 
For the $m_{4} \gg m_{3}$ model, inserting the expressions from Eq.~(\ref{eq:oscprobs43}) into (\ref{eq:fsdef}) shows that $f_{s}$ is energy dependent because of the constant terms in Eq.~(\ref{eq:oscprobs43}). For this reason the 90\% confidence level value of $f_{s}$ for this model is presented at $E_{\nu} = \unit[1.4]{GeV}$, the energy where the $\nu_{\mu}$ disappearance probability is a maximum.  The determination of the limit follows the procedure described above, but with the addition of selecting a value of $\theta_{24}$ for each test case as well. At 90\% confidence level $f_{s} < 0.52\,(0.55)$ for $E_{\nu} = \unit[1.4]{GeV}$ in this model.  Thus, in either model, approximately 50\% of the disappearing $\num$ can convert to $\nus$ at 90\% confidence level as long as the amount of $\nue$ appearance is less than the limit presented by the CHOOZ collaboration.

\section{Oscillations with Decay}\label{sec:Decay}
It was noted more than a decade ago that neutrino decay, as an alternative or companion process to neutrino oscillations, offers some capability for reproducing neutrino disappearance trends~\cite{ref:Decay}.  The model investigated here~\cite{ref:DecayMaltoni} includes neutrino oscillations occurring in parallel with neutrino decay.  Normal neutrino-mass ordering is assumed, and the mass eigenstates $\nu_1$,~$\nu_2$ are approximately degenerate, so that $m_3\gg m_2\approx m_1$. The heaviest neutrino-mass state $\nu_3$ is allowed to decay into an invisible final state. With these assumptions, and neglecting the small contributions from \nue~mixing, only the two neutrino flavor states $\nu_\mu$ and $\nu_\tau$,  and the corresponding mass states $\nu_2$ and $\nu_3$, are considered. The evolution of the neutrino flavor states is given by~\cite{ref:DecayMaltoni}:
%\begin{widetext}
\begin{eqnarray}\label{eq:decayEvo}
 i\frac{d\vec{\nu}}{dx}&=&\left[
\frac{\Delta m^2_{32}}{4E}\left(
\begin{array}{cc}
-\cos2\theta & \sin2\theta \\
\sin2\theta  & \cos2\theta
\end{array}
\right) \right.
\nonumber \\
&&-i\frac{m_3}{4\tau_3E}\left. \left(
\begin{array}{cc}
2\sin^2\theta & \sin2\theta \\
\sin2\theta   & 2\cos^2\theta
\end{array}
\right)\right]
\vec{\nu},
\end{eqnarray}
%\end{widetext}
where $\tau_3$ is the lifetime of the $\nu_3$ mass state and $\theta$ is the mixing angle governing oscillations between \numu~and \nutau. Solving Eq.~(\ref{eq:decayEvo}) one obtains probabilities for $\nu_\mu$ survival or decay:
%For a $\nu_\mu$ beam the evolution of the time-dependent \numu~state is given by:
%\begin{equation}
%  |\numu(t) \rangle=\cos\theta|\nu_1\rangle e^{i\tilde{p}_1\cdot\tilde{x}}+\sin\theta|\nu_2\rangle e^{\left(i\tilde{p}_2\cdot\tilde{x}-\frac{t}{2\tau_2}\right)}
%\end{equation}
%where $t$ is the proper time in the neutrino frame and $\tau_2$ is the lifetime of the second mass state. Electron neutrinos are taken not to participate in the oscillations. Following through the standard neutrino oscillation derivation one obtains probabilities for $\nu_\mu$ survival or decay:
% Is this the correct way to split a long equation over two lines?
\begin{eqnarray}
P_{\mu\mu} & = & \cos^4\theta+\sin^4\theta e^{-\frac{m_3L}{\tau_3E}}+\nonumber\\
          &   & 2\cos^2\theta\sin^2\theta e^{-\frac{m_3L}{2\tau_3E}}\cos\left(\frac{\Delta m^2_{32}L}{2E}\right)\\
P_{\rm decay} & = & \left(1-e^{-\frac{m_3L}{\tau_3E}}\right)\sin^2\theta. 
\label{eq:p_decay}
\end{eqnarray}
The limits $\tau_3\to\infty$ and $\Delta m^2_{32}\to0$ correspond to a pure oscillations or a pure decay scenario, respectively.

%% \begin{table}
%%   \begin{tabular}{c|c|c}
%%     mode & CC & NC \\
%%     \hline
%%     $\mu\to\mu$ & $c^4+s^4e^{-\alpha L/E}+2c^2s^2 e^{-\alpha L/2E}\cos{\Delta m^2L\over2E}$ & $1-(1-e^{-\alpha L/E})s^2$\\

%%     $\mu\to\tau$ & $c^2s^2(1+e^{-\alpha L/E}-2\cos{\Delta m^2L\over2E}e^{-\alpha L/2E})$ & 0 \\

%%     $\mu\to e$ & 0 & 0 \\

%%     $\mu\to s$ & $(1-e^{-\alpha L/E})s^2$ & $(1-e^{-\alpha L/E})s^2$ \\

%%     $e\to e$ & 1 & 1
%%   \end{tabular}
%%   \caption{Massively overdetailed decay plus oscillation formulae}
%% \end{table}

In a conventional neutrino oscillations scenario, the ratio of the predicted charged-current spectrum in the far-detector with the null-oscillation expectation displays the characteristic ``dip'' at the assumed $\Delta m^2_{32}$ value that is absent in the equivalent ratio computed for pure neutrino decay. Previously published results by MINOS using the charged-current far-detector spectrum support conventional oscillations and disfavor a scenario of pure neutrino decay at 3.7~standard deviations~\cite{ref:CCPRL}. In the present analysis, both the neutral-current and charged-current far-detector spectra shown in Figs.~\ref{fig:fd_spectrum}~and~\ref{fig:fd_spectrum_CC} are included in the fit. Consequently, additional sensitivity is gained with respect to previous analyses, since any neutrino decay into a noninteracting final state would also deplete the neutral-current spectrum according to Eq.~(\ref{eq:p_decay}). Additionally, the analysis is extended to the more general scenario combining oscillations and the decay model described above.

\begin{table}
  \begin{tabular}{|c*{3}{|c}|}
\hline
   Model & $\chi^2$/D.O.F. \bigstrut & $\alpha$~(GeV/km) & $\theta$\\
% Model 12
    \hline
    Osc. with Decay \bigstrut & 47.5/39 & $0.00^{+0.90}_{-0.0}\times10^{-3}$ & ${45.0}^{+10.83}_{-8.96}$    \\
    \hline
    Pure Decay  \bigstrut & 76.4/40 & $4.6^{+3.1}_{-2.3}\times10^{-3}$ & ${50.9}^{+39.1}_{-11.27}$ \\
    \hline
  \end{tabular}
  \caption{Best-fit points and uncertainty ranges obtained for the relevant parameters of the oscillation with decay model. The result obtained for the pure decay scenario, $\Delta m^2_{32}~\rightarrow0$, is also presented. Angles are shown in degrees.}
  \label{tbl:fit_decay}
\end{table}

% Figure~\ref{fig:fd_spectrum} shows the results of a fit to this model to the Far Detector data spectrum.

\begin{figure}
  \includegraphics[width=1.05\linewidth]{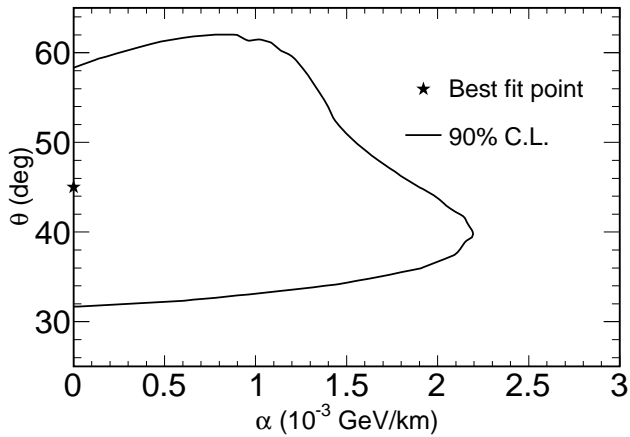}
  \caption[90\% C.L decay contour]{The best-fit point and 90\% C.L. contour for the two parameters of the neutrino oscillations-with-decay model, the neutrino mass-lifetime ratio $\alpha$ and the mixing angle $\theta$.}
  \label{fig:decay_contour}
\end{figure}

The best-fit values extracted  for $\theta$ and the parameter $\alpha\equiv m_3/\tau_3$ using this model are summarized in Table~\ref{tbl:fit_decay}. Figure~\ref{fig:decay_contour} shows the two-dimensional 90\% confidence interval obtained by the fit. The results are consistent with maximal mixing ($\theta=45^\circ$) and with no neutrino decay ($\alpha=0$). The best-fit value for $|\Delta m^2_{32}|$ is consistent with Ref.~\cite{ref:CCPRL}.

\begin{figure}
  \includegraphics[width=1.05\linewidth]{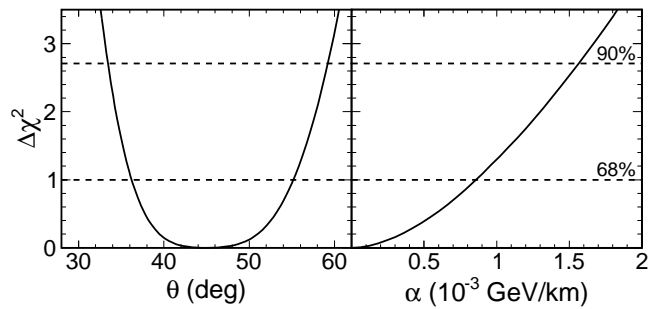}
  \caption{ Projections in $\Delta\chi^2$ for fit solutions for $\alpha$ and $\theta$ mixing angle for the oscillations-with-decay model. Parameter ranges allowed at 68\% and 90\% confidence levels lie below the corresponding dashed horizontal lines.}
  \label{fig:decay_projs}
\end{figure}
Figure \ref{fig:decay_projs} shows the one-dimensional $\Delta\chi^2$ projections for $\alpha$ and $\theta$, with other parameters marginalized. The 90\% confidence level limit found for the neutrino decay lifetime is $\tau_{3}/m_{3}~>~\unit[2.1\times10^{-12}]{\text{s/eV}}$.

A $\Delta\chi^2$ of 28.9 is obtained for the pure decay scenario. Thus, a pure neutrino decay model with null oscillations, as considered in Ref.~\cite{ref:CCPRL}, is disfavored at the level of 5.4~standard deviations.

% Turns out to be wrong by a factor 1e4
%% This result therefore rules out the fit point at ${\tau_3\over m_3}=2.6\times10^{-12}{\rm s/eV}$, (equivalent to $\alpha=1.28\times10^{-3}{\rm GeV/km}$) found in \cite{ref:GonzalezGarcia} with $>90\%$ confidence. TODO - exact $\Delta\chi^2$.

\section{Summary}
Searches for depletion or distortion in rate and visible energy spectra of neutral-current events recorded in the MINOS far-detector have been carried out for the purpose of detecting or constraining processes involving active-sterile neutrino mixing, as well as further restricting models including neutrino decay. The data exposure analyzed corresponds to \unit[$3.18\times10^{20}$]{protons on target} collected during the period of May 2005 to July 2007. 

 A total number of 388 neutral-current events were observed in the far-detector, whereas the expectation from standard three-flavor neutrino models is 377$\pm$19.4$\text{(stat.)}\pm$18.5$\text{(syst.)}$~events. The value for the statistic $R$ that gauges the agreement between the data and the expectation based on oscillations among the three active flavors is $R=1.04\pm0.08\text{(stat.)}\pm0.07\text{(syst.)}-0.10(\nue)$, which is consistent with no depletion of the active neutrino flux.

Joint fits to the observed neutral-current and charged-current energy spectra, assuming two different neutrino oscillation models that include an additional sterile neutrino flavor, yield the following 90\% confidence level limits on the oscillation parameters:
 $\theta_{34} < 38^\circ\,(56^\circ)$ for the model with $m_{4} = m_{1}$; $\theta_{24} < 10^\circ\,(11^\circ)$ and $\theta_{34} < 38^\circ\,(56^\circ)$ for the model with $m_{4}\gg m_{3}$.  The values in parentheses represent the results for maximally allowed \nue~appearance. These limits for the mixing angles between active and sterile neutrinos show that mixing between the active flavors dominates oscillations.  In fact, the fraction of active neutrinos that oscillate into a sterile species is constrained to be $f_{s} < 0.51\,(0.55)$ at the 90\% confidence level for the $m_{4} = m_{1}$ model and $f_{s} < 0.52\,(0.55)$ for  $E_{\nu}=\unit[1.4]{\text{GeV}}$ in the case of the model with $m_{4}\gg m_{3}$.  

Similar fits, assuming a two-flavor neutrino model in which oscillations may occur in parallel with decay into a sterile species, yield the best-fit values for the mass-lifetime ratio $\alpha=\unit[0.00^{+0.90}_{-0.0}\times10^{-3}]{\text{GeV/km}}$ and mixing angle $\theta={45.0^\circ}^{+10.83}_{-8.96}$. From these results, we extract a 90\% confidence level limit on the neutrino decay lifetime, \linebreak ${\tau_3}/{m_3}~>~\unit[2.1\times10^{-12}]{\text{s/eV}}$. The pure decay scenario ($\Delta m^2_{32}\rightarrow$0) is disfavored by 5.4~standard deviations in our data as an alternative explanation to neutrino oscillations.

\begin{acknowledgments}
We thank  S.~Parke and P.~Huber for useful discussions concerning oscillations between active and sterile neutrinos.  This work was supported by the U.S. Department of Energy, the U.S. National Science Foundation, the U.K. Science and Technology Facilities Council, the State and University of Minnesota, the Office of Special Accounts for Research Grants of the University of Athens, Greece, FAPESP (Funda\c{c}\~ao de Amparo a Pesquisa do Estado de S\~ao Paulo) and CNPq (Conselho Nacional de Desenvolvimento Cient\'ifico e Tecnol\'ogico) in Brazil.  We gratefully acknowledge the Minnesota Department of Natural Resources for their assistance and for allowing access to the facilities of the Soudan Underground Mine State Park.  We thank the crew of the Soudan Underground Physics laboratory for their tireless work in building and operating the MINOS far-detector.
\end{acknowledgments}

\end{document}